\documentclass[aps,prf]{revtex4}{

\usepackage{graphicx}

\usepackage{amsmath,amsfonts,bm, color,amssymb,float}
\usepackage{graphicx,epsfig,overpic,hyperref,ulem,caption,subcaption}
\usepackage{siunitx}

\newcommand{\refeq}[1]{Eq. [\ref{#1}]}
\newcommand{\reffig}[1]{Fig. \ref{#1}}

\newcommand{\col}[1]{{\color{black}{#1}}}
\newcommand{\colb}[1]{{\color{black}{#1}}}

\newcommand{\be}{\begin{equation}}
\newcommand{\ee}{\end{equation}}

\newcommand{\dif}{{\mathrm  d}}
\newcommand{\kT}{ k_BT}
\newcommand{\Km}{K_m}
\newcommand{\K}{{K_m}}

\newcommand{\Keff}{{\tilde\kappa}}

\newcommand{\sv}{{v}}
\newcommand{\bslip}{{b}}

\newcommand{\eps}{{\varepsilon}}
\newcommand{\out}{{\mathrm{-}}}
\newcommand{\ins}{{\mathrm{+}}}
\newcommand{\mem}{{\mathrm{s}}}

\newcommand{\hd}{{\mathrm{hd}}}
\newcommand{\br}{{\bf{r}}}

\newcommand{\bA}{{\bf A}}
\newcommand{\bB}{{\bf B}}
\newcommand{\bC}{{\bf C}}
\newcommand{\bD}{{\bf D}}

\newcommand{\bE}{{\bf E}}
\newcommand{\bF}{{\bf F}}
\newcommand{\bM}{{\bf{M}}}
\newcommand{\bI}{{\bf I}}
\newcommand{\bT}{{\bf T}}

\newcommand{\bU}{{\bf{V}}}
\newcommand{\bS}{{\bf y}}
\newcommand{\by}{{\bf y}}
\newcommand{\mACF}{{\bf R}}

\newcommand{\visrat}{{\chi}}
\newcommand{\tang}{{\mathrm{t}}}

\newcommand{\rhat}{{\bf \hat r}}

\newcommand{\that}{{\bm{\hat \theta}}}
\newcommand{\phihat}{{\bm{\hat \varphi}}}

\newcommand{\bt}{{\bf{t}}}
\newcommand{\bn}{{\bf{n}}}
\newcommand{\bev}{{\bf{e}}}
\newcommand{\bv}{{\bf v}}
\newcommand{\bu}{{\bf u}}

\newcommand{\im}{{\mathrm i}}

\newcommand{\jj}{{\ell}}

\newcommand{\half} {{\frac{1}{2}}}

\begin{document}

\title{Equilibrium fluctuations of  a quasi-spherical vesicle: role of the membrane dissipation}

\author{Petia M. Vlahovska\textit{$^{a}$} and Rony Granek\textit{$^{b}$}} 

\affiliation{\textit{$^{a}$}~Engineering Sciences and Applied Mathematics, Northwestern University, Evanston, IL 60208, USA; E-mail: petia.vlahovska@northwestern.edu, \textit{$^{b}$}~Avram and Stella Goldstein-Goren Department of Biotechnology Engineering, Dept. of Physics, and Ilse Katz Institute for Nanoscale Science and Technology, Ben-Gurion University of The Negev, Beer Sheva 84105, Israel }

\begin{abstract}

We theoretically investigate the thermally-driven curvature and lipid density fluctuations of a quasi-spherical vesicle, accounting for the dissipation due to monolayer viscosity and intermonolayer friction. The theory predicts that membrane curvature makes long-wavelength undulations sensitive to membrane viscosity  and speeds up the relaxation of the lipid density fluctuations. Implications for the dynamic roughness and Dynamic Structure Factor measurements of submicron liposomes on nano-second time scales 
are discussed. Specifically,  a clear stretched-exponential relaxation regime may not exist, in contrast to the behavior of  planar membranes for which an anomalous diffusion exponent of 2/3 has been predicted  [Zilman and Granek, Phys. Rev. Lett. (1996)].
\end{abstract}

\date{ \today}

\maketitle

\section{Introduction}

The biological function of membranes is closely tied to their flexibility \cite{Bassereau_2018}, commonly assessed through thermally driven undulations \cite{Brochard:1975,Watson:2011,Faizi:2020,Ipsen2019VesicleFluctuation,DimovaACIS:2014,Gupta:2021,Nagao:2023,Monzel:2016}.
 The canonical problem of 
 curvature fluctuations of a membrane (flickering) was considered in the pioneering work by Brochard and Lennon \cite{brochard.1975}. 
In this minimal model, the membrane is considered a structureless interface with bending rigidity $\kappa$;  an undulation with wavenumber $q$ of an initially planar membrane
is dissipated by the viscosity of the surrounding fluid $\eta$ and relaxes exponentially with a rate $\kappa q^3/4\eta$.

The undulation dynamics changes when the membrane  itself is curved
\cite{Fujitani1994,Olla:2000,Rochal:2005,Henle:2010, woodhouse_goldstein_2012,Rahimi:2013, Rahimi_thesis,Sigurdsson:2016,Vlahovska:2019,Sahu:2020},
  since 
 in-plane  and out-of-plane displacements couple \cite{Rochal:2005}. 
For a quasi-spherical, tensionless vesicle, whose shape is described in terms of fluctuating spherical harmonics modes 
 $r_s(\phi,\theta,t)=R\left(1+f(\phi,\theta,t)\right), \,\, f=\sum f_{\jj m}(t) Y_{\jj m}(\phi,\theta)$, 
 the relaxation rate of a mode amplitude $f_{\jj m}$ is predicted to be (in the case of a structureless membrane and same fluid inside and outside the vesicle) \cite{Fujitani1994,Olla:2000,Rochal:2005,Vlahovska:2019}
\begin{equation}
	\label{eqw}
			\hat\omega=\frac{\kappa}{\eta R^3}\frac{(\jj-1)\jj^2(\jj+1)^2(\jj+2)}{4 \jj^3+6 \jj^2-1+\left(4 \jj^2+4 \jj-8\right) \visrat_s}\,,
\end{equation}
where  {$\visrat_s=\eta_s/R\eta$} is a dimensionless membrane viscosity parameter, the ratio of the Saffman-Delbr\"uck length ($\eta_s/\eta$) to the vesicle radius $R$; \col{here $\eta_s$ and $\eta$ are the membrane and bulk viscosities}.
Setting $\visrat_s=0$ reduces \refeq{eqw} to the result for a non-viscous vesicle \cite{Milner-Safran:1987}.
For $\chi_s \gg 1$, a new regime is predicted to emerge in the relaxation spectrum of long-wavelength undulations, $1 \ll j \ll \chi_s$, in which dissipation is dominated by membrane viscosity and $\hat\omega(\jj)\simeq \frac{ \kappa }{4\chi_s\eta R^3}\, \jj^4$. \col{Provided this asymptotic regime is attained}, the corresponding anomalous diffusion (stretching) exponent $\alpha$, governing the membrane dynamic roughness and the dynamic structure factor (DSF), $S(k,t) \sim \exp!\left[-(\Gamma_k t)^{\alpha}\right]$--as measured in scattering experiments such as neutron spin echo (NSE) \cite{Nagao:2017}, dynamic light scattering \cite{Freyssingeas:1997}, X-ray photon correlation spectroscopy \cite{Falus:2005}, and fluctuation spectroscopy \cite{Betz:2012,Helfer:2001a}--becomes $\alpha = 1/2$, with $\Gamma_k \sim (k_B T)^2 k^4 R^2 / (\kappa \eta_s)$. This contrasts with the $\alpha = 2/3$ scaling and $\Gamma_k \sim (k_B T)^{3/2} k^3 / (\kappa^{1/2} \eta)$ predicted for large planar membranes \cite{Zilman-Granek:1996} and vesicles \cite{Granek:EPJE} in the limit of negligible membrane viscosity, where $\hat\omega(\jj)\simeq \frac{ \kappa }{4\eta R^3}\, \jj^3$ for $\jj\gg1$.
These predictions were recently experimentally confirmed in flickering of giant unilamellar vesicles \cite{Faizi:2024}, and also appear to apply to NSE data \cite{heinrich2025effect}.

The minimal, zero-thickness membrane model captures the membrane dynamics only to a limited extent because it neglects the bilayer architecture of the membrane. 
{More realistic approaches treat the membrane as an elastic thin plate, introducing corrections to account for finite thickness \cite{lipel}, or model it as a bilayer composed of two monolayers that can slide relative to each other.} \citep{Seifert-Langer:1993,Miao:2002,Krishnan:2018}.
Bending the membrane  stretches and compresses the outer and inner monolayers  \cite{Yeung:1995}. If monolayer slippage is allowed, the relaxation of the resulting lipid density difference, driven by monolayer compressibility and dissipated by lateral lipid flow and intermonolayer friction, has been shown to strongly affect the undulation dynamics of planar membranes at short times and wavelengths (submicron and nanoseconds)  \cite{Seifert-Langer:1993,Watson:2010,Watson:2011,Monroy:2009}, relevant to neutron spin echo and dynamic light scattering experiments \cite{Nagao:2023,freyssingeas1997quasi}. 
However,
even in that case, 
the relaxation rate of the bending mode 
of a planar membrane 
asymptotically remains of the same form as the Brochard and Lennon result with  bending rigidity $\kappa$ replaced by the unrelaxed bending modulus $\tilde\kappa=\kappa+2\Km d^2$ for  deformations measured at the bilayer midplane in the absence of any slip \cite{Seifert-Langer:1993, Watson:2011}; here $\Km$ is the monolayer compression modulus and $d$ is the monolayer thickness. The DSF stretching exponent of 2/3 also remains unchanged \cite{Watson:2010,Watson:2011}. \col{This behavior is often assumed to hold over the typical NSE time and length scales, effectively presuming that the system has reached its asymptotic relaxation regime \cite{Watson:2010, Watson:2011}. This raises the question: Is this assumption justified for liposomes? Is there an effect of the spherical geometry? }

 \refeq{eqw} shows that, for a quasi-spherical vesicle, the relaxation of long-wavelength undulations is dominated by membrane viscosity when the Saffman–Delbrück length exceeds the radius of curvature, i.e., $\chi_s \geq 1$. However, the zero-thickness model does not account for lipid density fluctuations.
Here, we extend the framework for the fluctuation dynamics of a quasi-spherical vesicle \cite{Miao:2002} to incorporate both intermonolayer friction and lipid density fluctuations, while also including membrane viscosity under the assumption that the monolayers behave as Newtonian fluids. The resulting theory provides a unified description of bilayer dynamics across experimentally relevant length and time scales.

\section{Problem statement}

\subsection{Membrane model}

The fluid bilayer membrane consists of two monolayers of amphiphilic molecules—typically lipids or polymers \citep{Seifert:1997,Discher:2006}. It exhibits a unique solid-fluid duality: it behaves as an elastic material in response to out-of-plane (bending) deformations, yet flows like a two-dimensional fluid under in-plane shear. The resistance to bending originates from the monolayers' finite thickness:
changes in curvature compress one monolayer while stretching the other --  {in addition to bending each monolayer} -- incurring  {an additional} elastic energy cost \cite{Evans-Skalak,Powers-Huber-Goldstein:2003}. In contrast, because the bilayer is held together by non-covalent bonds,  the lipids are free to rearrange and flow laterally within the monolayer..

\begin{figure}[h]
\centerline{\includegraphics[width=2.5in]{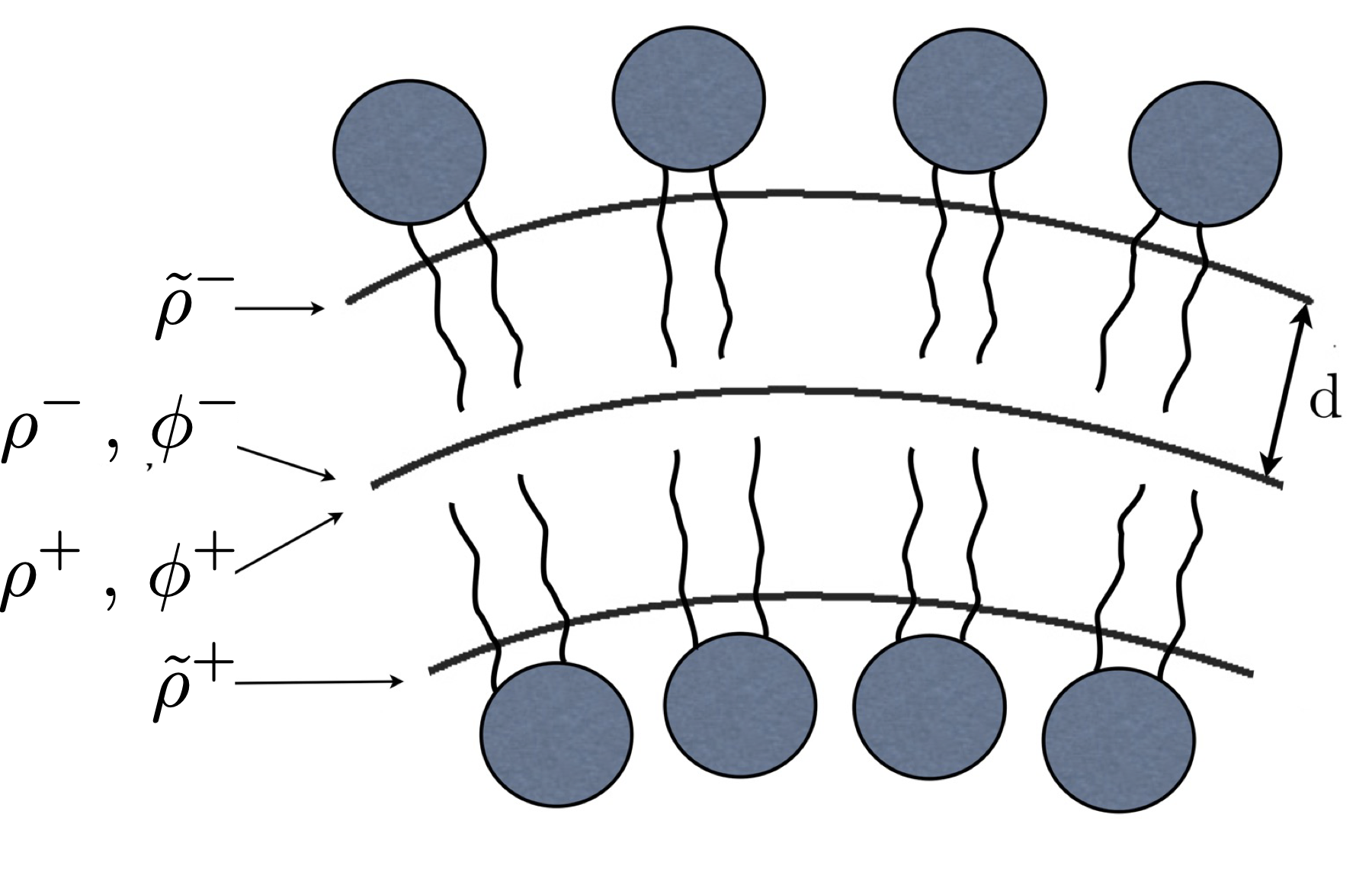}}
\caption{\footnotesize{{\bf{Structure of a lipid membrane formed by two identical monolayers}}. 
At the bilayer neutral surface bending and stretching are decoupled. 
${\rho}^\mp $ are the monolayers densities, $\tilde{\rho}^\mp $, projected onto the neutral surface. $\tilde{\rho}^\mp = \rho^\mp(1\pm 2dH)$, where $d$ is the monolayer thickness and $H$ is the mean curvature.
}}
\label{figSL}
\end{figure}

\subsubsection{Membrane elastic energy and forces}
We consider a membrane  composed of two  identical monolayers with  neutral surfaces that are distance $2d$ apart, see \reffig{figSL} \citep{Seifert-Langer:1993,Seifert:1997}. 
Compression and expansion of the monolayers 
adds to the bending energy 
\begin{equation}
\label{Energy1}
\begin{split}
  {\mathcal{H}}=&\frac{\kappa}{2}\int (2H)^{2}\dif A +\sigma_0\int  \dif A+\frac{\K}{2}\int \left[(\phi^{-}+2dH)^{2}+(\phi^{+}-2dH)^{2}\right] \dif A\,,
  \end{split}
\end{equation}
where the integration is over the vesicle area $A_0$ (defined by the neutral surface),  $\kappa$ is the bending modulus  {(twice the monolayer bending modulus)}, $H=-\half\nabla\cdot\bn$ is the mean curvature ($\bn$ being the outward pointing normal), 
$\sigma_0$ is the membrane tension, $\K$ is the monolayer compressibility modulus,  $\phi^\pm=\rho^\pm/\rho_0-1$, and the average equilibrium lipid density is $\rho_0=(\rho_0^++\rho_0^-)/2$, 
$\rho^\pm_0=N^\pm/A_0$. \refeq{Energy1} is rewritten as 
\begin{equation}
\label{FreeEnergy}
\begin{split}
\mathcal{H}
=&\frac{\Keff}{2} \int (2H)^{2}\dif A +\sigma_0\int \dif A+2 d \K \int H\left(\phi^- -\phi^+\right) \dif A+\frac{\K}{2} \int \left(\phi^+\right)^{2}\dif A+\frac{\K}{2} \int \left(\phi^-\right)^2\dif A,. 
 \end{split}
\end{equation}
The parameter $\Keff=\kappa + 2d^{2}\K$ is a renormalized bending rigidity and represents the bending
modulus for deformations measured at the bilayer midplane
in the absence of any density relaxation. 
The elastic terms have been expanded and grouped to give the third term which describes the energy cost associated with the coupling between changes in curvature and  local lipid densities  \cite{Miao:2002, Seifert-Langer:1993}.

A non-equilibrium membrane configuration, ${\bf{r}}$$_s$, exerts a force on the surrounding fluid,
$\bt=- \delta \cal{H}/\delta {\bf{r}}$$_s$  \cite{Seifert1999}.
The bending component for each monolayer is
\begin{equation}
\label{RestoringForcesB}
\bt_{\kappa}^\pm=\Keff\left[2H(H^{2}-K_G)+\nabla_s^2 H\right]\bn
\end{equation}
where $K_G$ is the Gaussian curvature, 
 $\nabla_s$ is the surface gradient operator, and $\nabla_s^2$ is the Laplace-Beltrami operator. 
 Each monolayer experiences tension $\sigma^\pm=\half \sigma_0-\K \phi^\pm-\frac{\K}{2}\left(\phi^\pm\right)^2$, which gives rise to tractions
 \begin{equation}
\label{RestoringForcesT}
\bt^\pm_{\sigma}=-\sigma^\pm\left(2H\right)\bn-\nabla_s\sigma^\pm
\end{equation}
 The curvature-lipid density coupling  gives rise to additional forces \cite{Miao:2002}
\begin{equation}
\label{RestoringForces}
\begin{split}
\bt_{c}^\pm=&
\mp 2 d \K\left(\half\nabla_s^2 \phi^\pm+(2H^{2}-K_G)\phi^\pm + 2H^{2}\right) \bn \mp 2d \K(1+\phi^{\pm})\nabla_{s}H\,.
\end{split}
\end{equation}

\subsubsection{Dissipation in the membrane}

The weak 
intermolecular interactions 
between the two monolayers allow the two monolayers to slide over each other  \citep{Evans_Yeung1992,Evans-Yeung:1994,Fournier:2015}, thereby making the tangential component of the velocity discontinuous. The friction due to the relative motion gives rise to 
 surface stresses on the monolayers facing the inner fluid ($\ins$) and the outer fluid ($\out$) 
\begin{equation}
    \bt^\pm_{\bslip} = \mp b(\bv_{\tang}^{\out}-\bv_{\tang}^{\ins})\,,
\end{equation}
where the parameter $b$ is the slip coefficient. Its magnitude varies greatly depending on the type of lipid; values of $b$ have been reported in the range $10^4 $ - $ 10^{9}\,\text{N.s}/\text{m}^3$ \citep{Shkulipa:2007,Merkel_Friction,Anthony:2022}. 

 Assuming the monolayers behave as Newtonian fluids, the viscous stresses 
 are given by the Boussinesq-Scriven equation \citep{Edwards-Brenner-Wasan:1991}
\begin{equation}
\label{BSstress} 
{\bm{\tau}}_{\sv}=
\nabla_s \cdot\left[\left(\eta_{md}-\eta_m\right) \left(\bI_s:\bD_s\right)\bI_s+2\eta_m \bD_s\right]\,,
\end{equation}
where $\eta_m$ and $\eta_{md}$ are  the 2D shear and dilatational monolayer viscosities; for a symmetric bilayer, the corresponding  membrane viscosities  are $\eta_s=2\eta_m$  and $\eta_d=2\eta_{md}$ .  $\bI_s=\bI-\bn\bn$, $\bI$ is the three-dimensional idemfactor, and the surface rate-of-strain tensor is
\begin{equation}
\bD_s=\frac{1}{2}\left[\nabla_s \bv_\mem \cdot \bI_s+\bI_s\cdot\left(\nabla_s\bv_\mem\right)^\dagger\right]\,,
\end{equation}
where the superscript ${\dagger}$ denotes transpose.
Note that in general the velocity of a deforming interface, $\bv_\mem$,  has both a normal and a tangential component to the interface, $\bv_\mem=v_n\bn+\bv_{\tang}$. 

The viscous interfacial tractions derived from \refeq{BSstress} \citep{Edwards-Brenner-Wasan:1991, Edwards_Wasan:1988,felderhof2006effect} have in general complicated expressions. These simplify in the case of  a  sphere 
\begin{equation}
\label{VstresS}
\begin{split}
\bt^{\pm}_{\sv}&=-\eta_{md}\frac{2}{R}(\nabla_s\cdot\bv_\mem^\pm) \rhat +(\eta_{md}+\eta_m) \nabla_s\nabla_s\cdot\bv_\mem^\pm+\eta_m\left(\rhat\times\nabla_s\left(\left(\nabla_s\times \bv_\mem^\pm\right)\cdot\rhat\right)-\frac{2}{R} \left(\nabla_s\bv_\mem^\pm\right)\cdot\rhat\right)
\end{split}
\end{equation}
where on a sphere with radius $R$, $\nabla_s\cdot\bv_\mem=\nabla_s\cdot\bv_\tang+2v_n/R$, $H=-1/R$  and $K_G=1/R^2$.

Data for membrane viscosities are scarce, and reported values vary greatly \cite{Faizi:2022,Fitzgerald:2023}.
Typical membrane shear viscosities for unsaturated lipids in the fluid phase  {obtained from micron-sized GUVs} are on the order of $\eta_s\sim 10^{-9}\,\mathrm{N.s/m}$ \citep{Faizi:2022}, although values obtained from molecular dynamic simulations  {of $\sim 10$ nm membrane patches} could be two orders of magnitude lower\cite{Zgosrki:2019,Fitzgerald:2023}.
Membranes in the liquid-ordered phase \cite{Faizi:2024} or those composed of polymers \cite{Dimova:1999,Faizi:2022} can exhibit significantly higher viscosities.
Measurements of the dilatational viscosity are even more limited \cite{Rumy:2006,Guzman:2022}, but the most recent reports indicate that dilatational and shear viscosities are of comparable magnitude \cite{Ponce:2017}.

\subsubsection{Dissipation in the bulk and fluid-membrane coupling}
Let us consider a vesicle suspended in a fluid with viscosity  $\eta^\out$ and enclosing fluid with viscosity $\eta^\ins$; both fluids are assumed incompressible and Newtonian. Fluid motion at the lenghtscales of a micron-sized vesicle  {and smaller} 
is in the overdamped regime, where inertia effects are negligible. Accordingly,  
the fluid velocity $\bv$ and pressure $p$ 
obey  the Stokes equations 
\begin{equation}
\label{Stokes equations}
\nabla\cdot\bT^{\pm}=-\nabla p^{\pm}+\eta^{\pm} \nabla^2\bv^{\pm}={0}\,,\quad \nabla\cdot\bv^{\pm}=0\,,
\end{equation}
where  $\bT$  is the bulk hydrodynamic stress tensor
\begin{equation}
\label{stress definition}
\bT= -p\bI+\eta\left[\nabla\bv+(\nabla\bv)^\dagger\right]\,.
\end{equation}
We will use  a  reference frame  centered at the vesicle. 
At the membrane, the normal component of the velocity is continuous (assuming an impermeable membrane), $\bv^\ins\cdot\bn=\bv^\out\cdot\bn\equiv \bv_\mem\cdot\bn$,  where $\bn$ is the outward pointing unit normal vector. The vesicle deformation is determined from 
the kinematic condition at the
interface 
\begin{equation}
\label{BC2}
\frac{\partial \br_s}{\partial t}=\bv_\mem\,,
\end{equation}
where $\br_s$ is the position vector of the neutral surface of the bilayer. 
 
 Mechanical equilibrium at each monolayer surface requires
\begin{equation}
\label{BCmon} 
\bt^{\hd,\pm}=\bt_\kappa^\pm+\bt_\sigma^\pm+\bt_c^\pm+\bt_{\sv}^\pm\mp b (\bv^\out_\tang-\bv_\tang^\ins)\,,
\end{equation}
where $\bt^{\hd,\pm}=\mp\bn\cdot\bT^\pm$ are the hydrodynamic tractions. This boundary condition can be expressed alternatively as
\begin{equation}
\label{BCmon2}
\bt^{\hd,-}-\bt^{\hd,+}=\bt_s^\Delta\,,\quad \bt^{\hd,-}+\bt^{\hd,+}=\bt_s^\Sigma\,,
\end{equation}
where $\bt_s^\Sigma$ and $\bt_s^\Delta$ are the sum and difference of the membrane tractions on the right-hand side of \refeq{BCmon}.

\subsubsection{Nondimensionalization}

Henceforth,  all variables are non-dimensionalized using the radius of a sphere with the same volume as the vesicle, the viscosity of the suspending (outer) fluid, and  the characteristic 
bending stress
$\tau_c=\Keff/R^3$.  
Accordingly, the time
scale is $t_c=\eta^\out/\tau_c$ and  the velocity scale is $V_c=R \tau_c/\eta^\out$. The membrane elastic and dissipative stresses give rise to the following dimensionless parameters
\begin{equation}
\begin{split}
&\alpha=\frac{\K R^2}{\Keff}\,,\quad \lambda=\frac{2\K d R}{\Keff}\,,\\
&\quad \visrat_s=\frac{\eta_s}{\eta^\out R}\,,\quad \visrat_d=\frac{\eta_d}{\eta^\out R}\,, \quad \beta=\frac{b R}{\eta^\out}\,, \quad \chi=\frac{\eta^\ins}{\eta^\out}
\end{split}
\end{equation}

\section{Dynamics of a quasi-spherical vesicle}
\label{sec3}

Let us consider a vesicle with total area $A_0$ of the neutral surface, and  enclosing fluid volume $V$. 
The asphericity (deflation) of the vesicle is characterized by a dimensionless excess
area
\begin{equation}
\label{excess area}
\Delta=A_0/R^2-4 \pi \,, 
\end{equation}
where the characteristic vesicle
size $R$ is defined by the radius of a sphere of the same volume, $R=\left(3V/4\pi\right)^{1/3}$.   We will consider a quasi- spherical vesicle for which $\Delta\ll 1$ and the excess area is stored in thermal undulations.  In this case, the 
instantaneous vesicle shape parametrized relative to a reference sphere 
centered  at the vesicle center of mass is 
\begin{equation}
\label{perturbation of shape}
r_s= R \left(1+ f(\theta, \varphi, t) \right)\,,
\end{equation}
where  $f \ll 1$ are the amplitude of the 
membrane undulations, whose magnitude is set by the competition of the thermal noise and resistance to bending. Considering that $<f^2>\simeq \kT/(12\pi \kappa)$ \cite{Granek:EPJE}, this implies that for  $\kappa\gtrsim\kT$ our linear analysis is reasonable. The $r$, $\theta$ and $\varphi$ are the radial distance,  polar and azimuthal angles in a spherical coordinate system.

The shape evolution is determined from the kinematic condition \refeq{BC2}
\begin{equation}
\label{eqFeps}
\frac{\partial f}{\partial t}=\bv_\mem\cdot\nabla f\,.
\end{equation}
 Since there is no exchange of lipids between the monolayers and between the monolayers and the bulk fluids, the total number of lipids in each monolayer is conserved
\begin{equation}
\label{eqLipidEq}
\frac{\partial \rho^\pm}{\partial t}+\nabla_s\cdot \left(\bv_\tang^\pm  \rho^\pm\right)+\rho^\pm(\nabla_s\cdot\bn)(\bv_\mem\cdot\bn)=0
\end{equation}
At equilibrium, the lipid density is $\rho^\pm_0/\rho_0=1+\phi_0^\pm= 1\mp \phi_0$, where 
\[
\phi_0=\frac{\rho^\out_0-\rho_0^\ins}{2\rho_0}=\frac{N^\out-N^\ins}{N^\out+N^\ins}
\]
is a measure of the equilibrium asymmetry in the lipid numbers of the monolayers: on a sphere equal lipid density of the monolayers implies different lipid numbers because the area of the outer monolayer is greater than the inner monolayer; for a sphere with radius $R$ and distance between the monolayers surfaces $2d$, $A_\out\approx A_{\ins}\left(1+2d/R\right)$, i.e., $\phi_0=2d/R$.

For small perturbations about the equilibrium lipid density,  $\rho^\pm/\rho_0=\rho^\pm_0/\rho_0+\phi^\pm$,
the lipid transport equation becomes
\begin{equation}
\label{eqLipidEq2}
\frac{1}{1\mp\phi_0}\frac{\partial \phi^\pm}{\partial t}+ \nabla_s\cdot \bv_\tang^\pm +(\nabla_s\cdot\bn)(\bv_\mem\cdot\bn)=0\,.
\end{equation}

\subsection{Solution}

 \subsubsection{Vesicle shape and energy}
 
 Due to the spherical geometry of the problem,  the vesicle shape and monolayer densities are expanded in spherical harmonics (see Appendix  \ref{Harmonics} for definitions)
\begin{equation}
\begin{split}
f(\theta, \varphi, t)=\sum_{\jj m} f_{\jj m} Y_{\jj m}\,,\quad \phi^\pm=\phi_{00}^\pm+\sum_{\jj m}  \phi^\pm_{\jj m}  Y_{\jj m}\,,
 \end{split}
\end{equation}
where the sum denotes $\sum_{\jj m} \equiv \sum_{\jj =2}^\infty\sum _{m=-\jj}^\jj$.  
It is more convenient to work 
with the two alternative fields, the lipid density difference between the monolayers and the  average lipid density 
\begin{equation}
\phi=\half\left(\phi^\out-\phi^\ins\right)\,,\quad \bar\phi=\half\left(\phi^\out+\phi^\ins\right)\,,
\end{equation}
which are expanded as follows
 \begin{equation}
 \phi=\phi_0+\sum_{\jj m} \psi_{\jj m}Y_{\jj m}\,,\quad \bar\phi=\sum_{\jj m} \xi_{\jj m}Y_{\jj m}
 \end{equation}
 Note that $\bar\phi_0=(\phi_0^\out+\phi^\ins_0)/2=0$ and $\phi_0=(\phi_0^\out-\phi_0^\ins)/2=(N^\out-N^\ins)/(N^\out+N^\ins)=\pm \phi_0^\mp$.
The energy \refeq{FreeEnergy} expanded in spherical harmonics, 
is 
\[
{\cal{H}}=\frac{\tilde\kappa}{2}\sum_{\jj m}\left(\begin{pmatrix}  {f}_{\jj m}\\ \psi_{\jj m}\end{pmatrix}^\dagger  \cdot\bE \cdot\begin{pmatrix}  f^*_{\jj m}\\ \psi^*_{\jj m}\end{pmatrix} +2\alpha \xi_{\jj m}^2\right) 
\]
where 
\begin{equation}
\label{mE}
\begin{split}
E_{11}=&(\jj-1)(\jj+2)\left(\jj(\jj+1)+\bar\sigma_0-\alpha \phi_0^2\right)\,,\\
E_{12}=&E_{21}=-(\jj-1)(\jj+2)\lambda\,,\\
E_{22}=&2\alpha\,.
\end{split}  
\end{equation}
$\bar\sigma_0=\sigma_0R^2/\Keff$ corresponds to the tension of a planar membrane, and for a sphere, 
 $\phi_0=2d/R$, if the lipid density is the same for both monolayers.
From the equipartition theorem 
\begin{equation}
\begin{split}
    \langle f_{\jj m}^2\rangle&=\frac{\kT}{\tilde\kappa} \frac{E_{22}}{E_{11}E_{22}-E_{12}E_{21} }=\frac{\kT}{ (\jj(\jj+1)-2)(\jj(\jj+1)\kappa+\sigma_0R^2)}
    \end{split}
\end{equation}
\begin{equation}
\begin{split}
    \langle \psi_{\jj m}^2\rangle&=\frac{\kT}{\tilde\kappa} \frac{E_{11}}{E_{11}E_{22}-E_{12}E_{21} } =\kT\frac{(\jj(\jj+1)\tilde\kappa+\sigma_0R^2-4\Km d^2)}{2\Km R^2(\jj(\jj+1)\kappa+\sigma_0R^2)}
     \end{split}
\end{equation}
\begin{equation}
\begin{split}
    \langle f_{\jj m}\psi^*_{\jj m}\rangle&=\frac{\kT}{\tilde\kappa} \frac{E_{12}}{E_{12}E_{21}-E_{11}E_{22} }=\kT\frac{d}{R (\jj(\jj+1)\kappa+\sigma_0R^2)}
     \end{split}
\end{equation}
 \begin{equation}
 \begin{split}
    \langle \xi_{\jj m}\xi^*_{\jj m}\rangle=\frac{\kT}{2\Km R^2}
     \end{split}
\end{equation}
 
 \subsubsection{Flow}
To solve for the flow,  we use the basis of fundamental solutions of the Stokes equations in a spherical geometry \cite{Vlahovska:2007,Vlahovska:2016a,Vlahovska:2019}, listed in Appendix~\ref{Ap:velocity basis},
\begin{equation}
\label{vel}
\bv^\out=\sum_{\jj mq} c^-_{\jj mq}\bu^-_{\jj mq}(\br)\,,\quad  \bv^{\ins}=\sum_{\jj mq} c^+_{\jj mq}\bu^+_{\jj mq}(\br)\,.
\end{equation}
$q$ takes values 0,1, and 2.   The functions
$\bu^{\pm}_{\jj mq}$ are vector solid spherical harmonics
 related to the harmonics in the  Lamb solution.  With respect to a sphere,
$\bu^{\pm}_{\jj m2}$
is radial, while $\bu^{\pm}_{\jj m0}$  and $\bu^{\pm}_{\jj m1}$ are tangential;
$\bu^{\pm}_{\jj m1}$ is surface-solenoidal ($\nabla_s \cdot
\bu^{\pm}_{\jj m1}=0$).
 The velocity coefficients $c^\pm_{\jj mq}$ are
determined from the condition for velocity continuity and the stress balance.

\subsubsection{Shape and lipid density evolution}

The  shape evolution equation, \refeq{eqFeps}, to a leading order is 
\begin{equation}
\label{eveqLeadingOrder}
 \dot f_{\jj m}=c^+_{\jj m2}=c^-_{\jj m2}\,.
\end{equation}
Here the dot denotes a time derivative. 
The redistribution of the lipids \refeq{eqLipidEq2} yields
\begin{equation}
\label{eqL}
\begin{split}
\dot  \psi_{\jj m}&=-2\phi_0 \dot f_{\jj m}+\half{\jj\left(\jj+1\right)}\left(\left(1+\phi_0\right)c^\out_{\jj m0}-\left(1-\phi_0\right)c^\ins_{\jj m0}\right)\,,\\
 \dot \xi_{\jj m}&=-2 \dot f_{\jj m}+\half{\jj\left(\jj+1\right)}\left(\left(1+\phi_0\right)c^\out_{\jj m0}-\left(1-\phi_0\right)c^\ins_{\jj m0}\right)
 \end{split}
\end{equation}
We neglect lipid flip-flop and thus the total number of lipids in each monolayer is constant.
\col{An examination of the characteristic  time scales corresponding to relaxation of perturbations in the curvature and monolayer density  -- bending limited by solvent viscous dissipation ($t_\kappa$), and  monolayer compression/expansion limited by either  sliding friction between the monolayers ($t_b$) or lateral lipid flow ($t_{K}$) -- shows that}
\begin{equation}
	t_\kappa = \frac{\eta^\out R^3}{\kappa}\sim 12 \mathrm{s}\,,\quad t_b = \frac{b R^2}{\K}\sim 0.05 {\mathrm s} \,,\quad t_{\K}=\frac{\eta^\out R}{\K}\sim 5\times 10^{-7} {\mathrm s} \,,
\end{equation}
for a GUV with radius $R = 10^{-5}\text{ m}$, where $\eta^\out = 0.001\text{ N.s}/\text{m}^2$, $\kappa = 20\kT$, $\K = 0.02\text{ N}/\text{m}$, and $b = 10^7 \text{ N.s}/\text{m}^3$ \cite{Seifert-Langer:1993}. Even for liposomes with submicron radius, the time scale for the average density relaxation remains much faster than the relaxation of the bending and lipid density fluctuations (for example, for a $R=50$nm,  $t_\kappa \sim t_b \sim 1  {\mathrm \mu s} \,,t_{\K}\sim 2 {\mathrm ns} $).  The large separation of time scales allows us to set \cite{Goldstein:1996, Miao:2002},
\begin{equation}
\label{LocalAreaConstraint}
	\dot \xi_{lm} = 0\,.
\end{equation}
This condition is analogous to the  ``local area incompressibility'' constraint used in the zero-thickness model. It means that local changes in the average of the projected densities relax faster than changes in shape  or local density difference \cite{Goldstein:1996}. As pointed out in \cite{Miao:2002} this constraint is not to be directly compared to incompressibility of the bulk fluids because only one mode of monolayer density relaxation, $\xi_{lm}$, can be eliminated by invoking it ($\psi_{lm}$ remains). This added condition is treated as an extra constraint in the system which has to be explicitly enforced. 

Setting $\dot \xi_{\jj m}=0$ in \refeq{eqL}, solving for $c^-_{\jj m 0}$ yields 
\begin{equation}
\label{c0}
c^+_{\jj m 0}=\frac{1}{1-\phi_0}\left(\frac{4 f_{\ell m}}{{\ell(\ell+1)}}-c^-_{\jj m 0}(1+ \phi_0)\right)
\end{equation}
Inserting this result in the expression for $\dot \psi_{\jj m}$ leads to 
\begin{equation}
\label{eqPsi}
 \dot  \psi_{\jj m}=-\left(1+\phi_0\right)\left(2 \dot f_{\jj m}+{\jj\left(\jj+1\right)}c^\out_{\jj m0}\right)
\end{equation}
\subsubsection{Interfacial stresses}
\label{sec:memstres}
For a quasi-spherical vesicle, we  consider small deviations from equilibrium, $H=-1-C$ and $K=1+2C$, see \refeq{Hcurv} and \refeq{Kcurv}, $\phi^\pm=\phi_0^\pm+\tilde\phi^\pm$, and the $\sigma^\pm=\bar \sigma^\pm+\tilde \sigma^\pm$
where
\[
\bar \sigma^\pm=\frac{\sigma_0}{2}-\alpha \phi_0^\pm -\frac{\alpha}{2} \left(\phi_0^\pm \right)^2\,,\quad \tilde \sigma^\pm=-\alpha \tilde \phi^\pm-\alpha \phi_0^\pm\tilde \phi^\pm
\]
To the linear order in the shape  and lipid density  deviations from equilibrium, the radial components of the membrane elastic tractions \refeq{RestoringForcesB} and \refeq{RestoringForces} are
\begin{equation}
\begin{split}
&\bt^\pm_\kappa\cdot \bn \approx -2\nabla^2_sC  \,,\quad \bt^\pm_\sigma\cdot \bn \approx 2\bar \sigma  +2\tilde\sigma  \,,\\
&\bt^\pm_c\cdot \bn\approx\mp\lambda\left(\half \nabla^2_s\tilde \phi^\pm +\tilde \phi^\pm+2\phi_0^\pm C+4C\right)
\end{split}
 \end{equation}
 where constant terms are omitted since these are balanced by hydrostatic pressure.
 The tangential tractions are 
 \begin{equation}
 \label{ttrac}
\begin{split}
\bt_t^\pm=-\nabla_s\sigma^\pm\mp \lambda(1+\phi^\pm)\nabla_s H\approx (1+\phi_0^\pm)\left(\alpha\nabla_s\tilde\phi^\pm\pm \lambda\nabla_sC\right)
 \end{split}
 \end{equation}
In terms of spherical harmonics, the interfacial stresses are written as
\begin{equation}
\bt\cdot\rhat=\tau_{\jj m2} Y_{\jj m}\,,\quad \bt_t=\tau_{\jj m0}\nabla_s Y_{\jj m}\,.
\end{equation}
The tangential tractions derived from  \refeq{ttrac} are
\begin{equation}
\label{strac}
\begin{split}
\tau^{\Sigma}_{jm0}=&\tau^{\out}_{jm0}+\tau^{\ins}_{jm0}=2\alpha\left(\xi_{\jj m}+\phi_0 \psi_{\jj m}\right)-\lambda \phi_0 \left(\jj+2\right)\left(\jj-1\right)f_{\jj m}\\
\tau^{\Delta}_{jm0}=&\tau^{\out}_{jm0}-\tau^{\ins}_{jm0}=2\alpha\left(\psi_{jm}+\phi_0 \xi_{\jj m}\right)-\lambda \left(\jj+2\right)\left(\jj-1\right)f_{\jj m}
\end{split}
\end{equation}
 
For the normal stress balance, only the sum of  the elastic tractions matters
\begin{equation}
\begin{split}
\tau^{ \Sigma}_{jm2}=\tau^{\out}_{jm2}+\tau^{\ins}_{jm2}&=
\left(\jj+2\right)\left(\jj-1\right)\left(\jj\left(\jj+1\right)+\tau_0\right)f_{\jj m}-4\alpha \xi_{\jj m}-\left(4\alpha \phi_0+\lambda\left(\jj+2\right)\left(\jj-1\right) \right)\psi_{\jj m}
\end{split}
\end{equation}
where $\tau_0=\sigma_0R^2/\Keff-\alpha \phi_0^2+2\lambda \phi_0$. 
The traction associated with the bilayer friction, after using \refeq{c0} and \refeq{eqL} to express $c^+_{\ell m 0}$ in terms of $\dot\psi_{\ell m}$, is
\begin{equation}
\tau^b_{jm0}=\tau^{b,\out}_{jm0}-\tau^{b,\ins}_{jm0}=2\beta\left(c_{jm0}^--c_{jm0}^+\right)=-\frac{4\beta}{\jj(\jj+1)\left(1-\phi_0^2\right)}\dot\psi_{\jj m}\,.
\end{equation}
 Note that $\tau^{b,\out}_{jm0}+\tau^{b,\ins}_{jm0}=0$.

In the case of a  viscous area-compressible interface, the stresses  obtained from \refeq{VstresS} are \citep{Schwalbe4} 
\begin{equation}
\label{Estress}
\begin{split}
\tau^{\sv,\pm}_{\jj m0} &=\half \visrat_s {(\jj-1)(\jj+2)}c^\pm_{\jj m0}+\half \visrat_d \left(-2c^\pm_{\jj m2}+{\jj(\jj+1)} c^\pm_{\jj m0}\right)\,,\\ 
\tau^{\sv,\pm}_{\jj m2} &=\visrat_d \left(-2c^\pm_{jm2}+\jj (\jj+1)c^\pm_{\jj m0}\right)\,.
\end{split}
\end{equation}
Using the condition for the area incompressibility \refeq{eqPsi} and \refeq{eveqLeadingOrder} yields for the viscous stresses
\begin{equation}
\label{SVstress}
\begin{split}
\tau^{\sv, \Delta}_{\jj m0}= &\frac{\jj(\jj+1) (\visrat_s+\visrat_d)-2\visrat_s}{1-\phi_0^2}\dot\psi_{\jj m}\,, \\
\tau^{\sv,\Sigma}_{\jj m0}= &\frac{(2 \chi_s - \jj (1 + \jj) (\chi_d + \chi_s))\phi_0 \dot\psi_{\jj m} + 
 2 (-2 + \jj + \jj^2) \chi_s(1 - \phi_0^2) \dot f_{\jj m}}{\left(1-\phi_0^2\right){\jj(\jj+1)}}\,,\\
\tau^{\sv, \Sigma}_{\jj m2} &=- \frac{2\visrat_d \phi_0}{1-\phi_0^2}\dot\psi_{\jj m}\,,
\end{split}
\end{equation}

The condition for incompressibility, implies that $\xi_{\jj m}$ would adjust to keep $\dot\xi_{\jj m}=0$;  $\alpha\xi_{\jj m}$ acts as a tension counteracting imposed stresses to keep the area elements on the neutral surface from expanding/compressing. The ``tension'' $\alpha\xi_{\jj m}$ has two contributions: one balancing the elastic membrane stresses and one balancing the viscous stresses.  The elastic contribution is determined from setting $\tau^{\Sigma}_{jm0}=0$ in \refeq{strac}. Solving for $\xi_{jm}$  yields
\begin{equation}
    \xi^{el}=-\phi_0\psi_{\jj m}+\frac{(-2 +\jj(\jj+1))\lambda \phi_0}{2 \alpha}f_{\jj m}
\end{equation}
Inserting in the expression for the tangnetial $\tau^{\Delta}_{jm0}$ and normal elastic  stresses $\tau^{\Sigma}_{jm0}$  leads to modified  elastic stresses
\begin{equation}
\begin{split}
\tau^{\Delta}_{jm0}&=\left(1-\phi_0^2\right) \left(2\alpha \psi_{jm} -\lambda \left(\jj+2\right)\left(\jj-1\right)f_{\jj m}\right)\,,\\
\tau^{\Sigma}_{jm2}&=\left(\jj+2\right)\left(\jj-1\right)\left(\left(\jj\left(\jj+1\right)+\tau_0-2\lambda\phi_0\right)f_{\jj m}-\lambda \psi_{\jj m}\right)\,.
\end{split}
\end{equation}
The tension arising from viscous forces is obtained from the tangential stress balance 
\begin{equation}
(\tau^{\hd,-}_{jm0}+\tau^{\hd,+}_{jm0})-\tau^{\sv, \Sigma}_{jm0}=-2\alpha \xi^{v}_{jm}\,.
\end{equation}
The hydrodynamic tractions are listed in Appendix~\ref{Ap:velocity basis}. Solving for the average density ``strain" yields
\begin{equation}
\begin{split}
\xi^{v}_{jm}&=-[{\alpha \jj(\jj+1)\left(1-\phi_0^2\right)}]^{-1}\\
&\times \left(-(2 + \jj +\chi( \jj-1) + 2 (-2 + \jj + \jj^2) \chi_s) \left(1 -\phi_0^2\right) \dot f_{\jj m} \right.\\
&\left. +\left(\jj (1 + \jj) \chi_d \phi_0 + (-2 + \jj + \jj^2) \chi_s \phi_0 + (1 + 2 \jj) (-1 + \chi +  (1+\chi) \phi_0)\right)\dot\psi_{\jj m}\right)
    \end{split}
\end{equation}

\subsubsection{Shape and lipid density evolution equations}

The stress boundary conditions yield the system of equations describing the shape and density dynamics.
\begin{equation}
\label{AE}
\begin{split}
(\tau^{\hd,-}_{jm2}+\tau^{\hd,+}_{jm2})-\tau^{\sv, \Sigma}_{jm2}+4\alpha \xi^v_{\jj m}=\tau^\Sigma_{jm2}\\
(\tau^{\hd,-}_{jm0}-\tau^{\hd,+}_{jm0})-\tau^{\sv,\Delta}_{jm0}+\tau^b_{jm0}-2\alpha \phi_0 \xi^v_{\jj m}=\tau^{\Delta}_{jm0}
\end{split}
\end{equation}
The full expressions, listed in the Appendix, \refeq{evoleqsFull}, are well approximated by
\begin{equation}
\label{evoleqsSL}
\bM\cdot \begin{pmatrix}  \dot{f}_{\jj m}\\ \dot{\psi}_{\jj m}\end{pmatrix}= -\bE\cdot \begin{pmatrix}  {f}_{\jj m}\\ \psi_{\jj m}\end{pmatrix}
\end{equation}
\begin{equation}
\label{mM}
\begin{split}
M_{11}=&
\frac{9+(-5+3\jj^2+2\jj^3)(\visrat+1)+4(-2+\jj+\jj^2)\visrat_s}{\jj(\jj+1)}\\
&\approx 2\jj(\visrat+1)+4\visrat_s\quad \mbox{for  }\jj\gg 1\\
M_{12}=&
M_{21}=0\\
M_{22}=&\frac{4\beta+2\jj+1+(2\jj+1)\visrat +\jj(\jj+1)\visrat_d +(-2+\jj+\jj^2)\visrat_s}{\jj(\jj+1)}\\
&\approx 4\beta/\jj^2+2\visrat/\jj+(\visrat_d+\visrat_s) \quad \mbox{for  }\jj\gg 1
\end{split}
\end{equation}
{Note that in this approximation {\bf M}$^{-1}$ is the Onsager matrix,
implying that the cross Onsager coefficients vanish. }

\col{Noteworthy, our model is restricted to wavelengths longer than the bilayer thickness. Specifically, the high mode cutoff is $\ell_{max}=\pi R/(2d)$, 
and the correction due to membrane thickness introduced in Ref. \cite{lipel} is negligible.
Moreover, since the shortest relaxation time $t_0\equiv 1/\hat \omega(\ell=\ell_{max})$ is typically at least one order of magnitude shorter than a nanosecond (i.e., $\sim 0.01-0.1$ ns), this finite thickness correction\cite{lipel} leads to a negligible contribution to the dynamic roughness (MSD) at times $t\gtrsim 1$ ns relevant for the application to the NSE experiments. }

{\subsection{Time-correlation functions}}

The exponential solution to \refeq{evoleqsSL} is found by
first diagonalizing \cite{Watson:2011}
\[
\bC=\bU \begin{pmatrix} \gamma_1 & 0\\ 0 & \gamma_2\end{pmatrix}\bU^{-1}
\]
where the matrix $\bC=\bM^{-1}\cdot\bE$, and  $\bU$ is the matrix of the eigenvectors of $\bC$. The relaxation rates correspond to the eigenvalues of the matrix $\bC$:
\begin{equation}
\label{decayrates}
\gamma_{1,2}=\frac{1}{2}\left(C_{11}+C_{22}\mp\sqrt{\left(C_{11}-C_{22}\right)^2+4 C_{12}C_{21}}\right)
\end{equation}
The general expressions for the decay rates $\gamma_1$ and $\gamma_2$ are complicated, 
but at high wavenumbers, $\jj\gg 1$
\[
\gamma_1\sim \frac{2\kappa\alpha}{\tilde\kappa\visrat_s}\,,\quad\gamma_2\sim\frac{\jj^3}{4} \quad ( \mbox {if } \visrat_s\ll1)\,,\mbox{ or } \gamma_2\sim\frac{\jj^4}{4\visrat_s} \quad  (\mbox{ if } \visrat_s\gg 1)\,.
\]
At low wavenumbers, 
$
\gamma_1=\omega(\jj, \kappa,\chi_s)\,,\quad\gamma_2\sim\frac{\alpha \jj^2}{2\beta} \,,
$
where 
\begin{equation}
\label{eqSw}
\omega
=\frac{\kappa}{\tilde\kappa}\left(\frac{(\jj-1)\jj(\jj+1)(\jj+2)\left(\jj(\jj+1)+\bar\sigma \right)}{\left(4 \jj^2+4 \jj-8\right) \visrat_s+\left(2 \jj^3+3 \jj^2-5\right) \left(\visrat-1\right)+4 \jj^3+6 \jj^2-1}\right)
\end{equation}
For tensionless membrane and same fluids inside and outside the vesicle, $\visrat=1$, the above expression reduces to \refeq{eqw} in dimensional form, $\hat\omega=\omega/ t_\Keff$.
The time correlation functions are the elements of the matrix
\[
\mACF=\bU \begin{pmatrix}  e^{-\gamma_1 t} & 0\\ 0 & e^{-\gamma_2 t}\end{pmatrix}\bU^{-1}\bE^{-1}
\]
Specifically
\begin{equation}
\langle f_{\jj m}(t)f_{\jj m}^*(0)\rangle =\mACF_{11}= \langle f_{\jj m}^2\rangle \left(Q_{11} e^{-\gamma_1 t}+(1-Q_{11}) e^{-\gamma_2 t}\right)
\end{equation}
with 
\begin{equation}
Q_{11}=\frac{\gamma_2-\omega}{\gamma_2-\gamma_1}\,,\quad \omega=C_{11}-\frac{E_{21}C_{12} }{E_{22}}\,.
\end{equation}
where the tension is the equilibrium one.
The density-density correlations
\begin{equation}
\langle \psi_{\jj m}(t)\psi_{\jj m}^*(0)\rangle =\mACF_{22}= \langle \psi_{\jj m}^2\rangle \left(Q_{22} e^{-\gamma_1 t}+(1-Q_{22}) e^{-\gamma_2 t}\right)
\end{equation}
with 
\begin{equation}
Q_{22}=\frac{\gamma_2-\omega_d}{\gamma_2-\gamma_1}\,,\quad \omega_d=C_{22}-\frac{E_{12}C_{21} }{E_{11}}\,.
\end{equation}
The shape-density correlations
\begin{equation}
\begin{split}
\langle f_{\jj m}(0)\psi_{\jj m}^*(t)\rangle&=\langle \psi_{\jj m}(0)f_{\jj m}^*(t)\rangle=\mACF_{12}\\
&= \langle f_{\jj m}\psi_{\jj m}^*\rangle \left(Q_{12} e^{-\gamma_1 t}+(1-Q_{12}) e^{-\gamma_2 t}\right)\\
\end{split}
\end{equation}
with 
\begin{equation}
\begin{split}
Q_{12}=&\frac{\gamma_2}{\gamma_2-\gamma_1}
\end{split}
\end{equation}

{\subsection{Mean Square Displacement and Dynamic Structure Factor}}
{{\subsubsection{General}}}

Scattering techniques, such as neutron spin echo
\cite{Nagao:2017}, dynamic light scattering \cite{Freyssingeas:1997}, X-ray photon correlation spectroscopy \cite{Falus:2005} and some flickering experiments   \cite{Betz:2012,Helfer:2001a} measure
DSF, $S(k,t)$, that is controlled by the single-point membrane mean square displacement (MSD), $\langle (\Delta h(t))^{2}\rangle$, and essentially captured by \cite{Zilman-Granek:1996, Watson:2011, Zilman-Granek:2002}
\begin{equation}
    S(k,t)\sim \text{Exp}[-\frac{k^2}{2}\langle (\Delta h(t))^{2}\rangle]\,,
\end{equation} 
{where $k$ is the scattering wavenumber (not to be confused with the undulation wavenumber $q=\ell/R$).} The dimensionless membrane segment MSD at an arbitrary 3D angle $\Omega=(\theta,\phi)$, 	$\langle (\Delta f(t))^{2}\rangle\equiv \langle (f(\Omega,t)-f(\Omega,0))^{2}\rangle$, is given by 
\begin{equation}
\begin{split}
	\langle (\Delta f(t))^{2}\rangle =
	\frac{1}{2\pi} \sum_{\jj =2}^{\jj_{\max}} 
	  (2\jj +1) \left(\langle |f_{\jj m}|^2\rangle-\langle f_{\jj m}(t)f_{\jj m}^*(0)\rangle\right)\,.
	\end{split}
	\label{segment-MSD}
\end{equation}
The MSD with physical dimensions is given by $\langle (\Delta h(t))^{2}\rangle\equiv R^2 \langle (\Delta f(t))^{2}\rangle$.

{{\subsubsection{Case of relaxed lipid density}}}

At times when the lipid density is relaxed \cite{Faizi:2024}, and assuming vanishing tension
\begin{equation}
		\langle (\Delta h(t))^{2}\rangle\equiv R^2 \langle (\Delta f(t))^{2}\rangle\approx \begin{cases}
		\frac{\Gamma[1/3]}{2\pi 4^{2/3}} \frac{\kT}{\eta^{2/3}\kappa^{1/3} }\, t^{2/3} & t_0 \ll {t \ll t^{*}}\\
		\frac{1}{4\sqrt{\pi}} \frac{\kT R}{\sqrt{\kappa \eta_s}}\, t^{1/2} &  t^{*}\ll t\ll \tau_R
	\end{cases}
	\label{MSD-asymptotes}
\end{equation}
where $t_0$ and $\tau_R$ are the shortest and longest relaxation times (respectively), $t_0\equiv 1/\hat \omega(\jj=\jj_{max})$ and $\tau_R \equiv 1/\hat\omega(\jj=2)$, and the crossover time $t^{*}\approx \tau_R \chi_s^{-4}$.
It follows that the scattering from vesicles in this time range and large scattering wavenumbers, $k R\gg 1$, would still exhibit a stretched exponential DSF
\begin{equation}
	S(k,t)\approx S(k)\times \begin{cases}
		\text{Exp}[-(\Gamma^{ZG}_k t)^{2/3}] & t_0 \ll t \ll t^{*}\\
		\text{Exp}[-(\Gamma^{VG}_k t)^{1/2}] &  t^{*}\ll t\ll \tau_R
	\end{cases}
	\label{DSF-asymptotes}
\end{equation}
where $\Gamma^{ZG}_k\simeq \frac{}{} (\kT)^{3/2}k^3/\kappa^{1/2}\eta$
 is the ZG relaxation rate 
and the new, membrane-viscosity-controlled, relaxation rate is given by
\begin{equation}
	\Gamma^{VG}_k\simeq \frac{(\kT)^2 R^2}{64\pi \kappa \eta_s} k^4
\end{equation}
Given that $t^{*}$ can be extremely short for viscous membrane vesicles with $R\sim 20-50$ nm,
it is quite possible that the entire NSE time window is controlled by membrane viscosity. Note that this prediction for the DSF does not account for finite-size effects arising from scattering by a spherical shell \cite{Granek:EPJE}. In addition, since 
$\Gamma_{k}^{VG}$ depends on $R$, polydispersity is expected to modify the decay profile. Finally, lipid density relaxation also affects the DSF decay and limits the validity of the predicted asymptotics. We explore the coupled effects of curvature and density fluctuations in the next section.

\section{Results}

\begin{figure}[h]
\includegraphics[width=0.5\columnwidth]{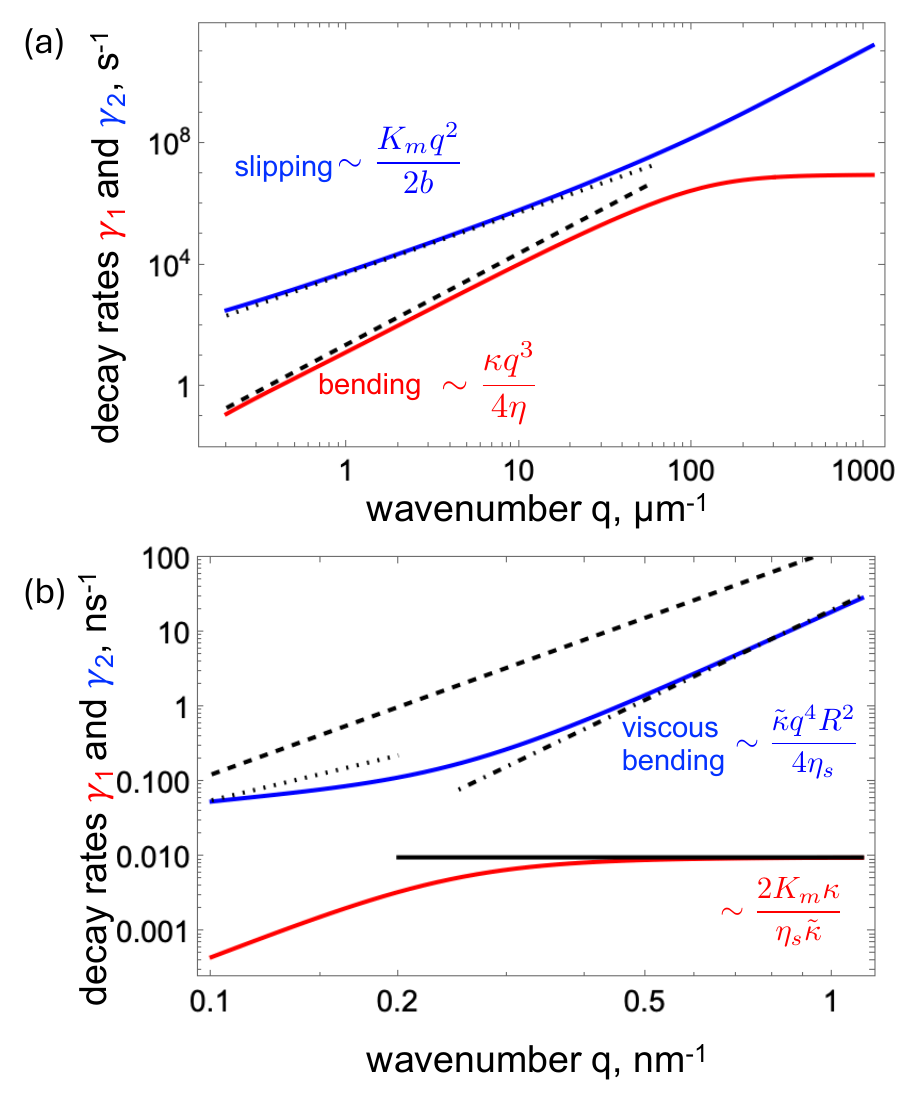}
\caption{\footnotesize{ Relaxation rates for vesicles made of DMPC against undulation wavenumber  $q=\ell/R$: (a) a GUV with radius  $R=10$ $\mu$m, $\chi_s=0.25$. (b) a SUV with radius  $R=20$ nm, $\chi_s=125$. Material parameters from Ref. \cite{Watson:2011}, $\kappa=13.6\,\kT$, $K_m=0.117$ N/m, $d=1.4$nm, $b=10^7$ N.s/m$^3$, $\eta_s=2.5\times 10^{-9}$ N.s/m, $\chi_d=0$, and bulk viscosity $\eta^\ins=\eta^\out\equiv\eta=10^{-3}$ N.s/m$^2$.
 The black long-dashed line corresponds to the asymptotic behavior of the bending mode from the Seifert-Langer theory for a planar membrane, $\tilde \kappa q^3/4\eta$. The  short-dashed line is the slipping mode, $\Km q^2/2b$. The solid black line is the viscous mode, $2\Km \kappa/\eta_s \Keff$. The dot-dashed line is the new asymptotic behavior,  $\tilde \kappa q^4 R^2/4\eta_s$, for relaxation controlled by dissipation by membrane viscosity. 
}}
\label{fig1}
\end{figure}

In this section, we explore the curvature and lipid density fluctuations of a quasi-spherical vesicle. Figure \ref{fig1} illustrates the relaxation dynamics for vesicles made of a typical lipid. The decay rates are computed from \refeq{decayrates} using the material properties listed in Ref. \cite{Watson:2011}. 

\begin{figure}[h]
\includegraphics[width=0.5\columnwidth]{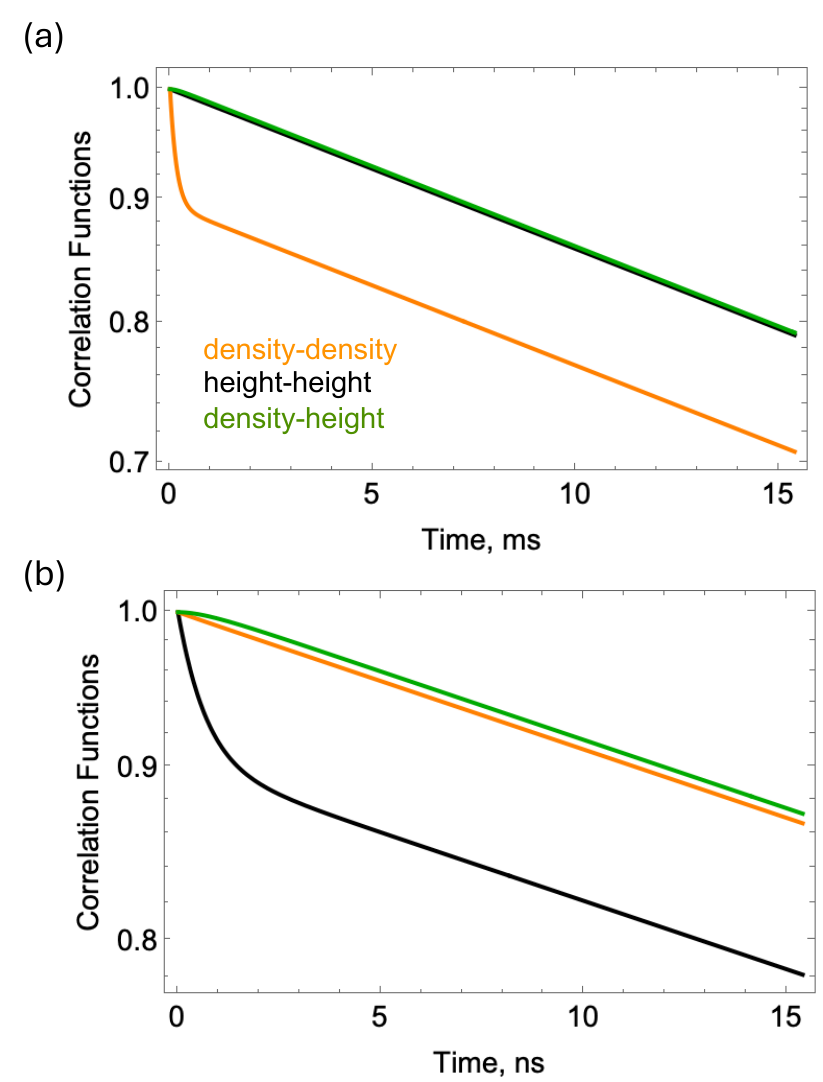}
\caption{\footnotesize{Correlation functions for the curvature and density fluctuations of mode $q=10$.
{(a) and (b) correspond to the same parameters as in Figs. 2(a) (GUV) and 2(b) (SUV), respectively.}
\label{fig2}
}}
\end{figure}

The lipid bilayer has a low membrane viscosity; accordingly, for a GUV, $\visrat_s \ll 1$. As shown in \reffig{fig1}a, the fast ``slipping'' mode ($\gamma_2$) decays on millisecond timescale or faster, while the slow ``bending'' mode ($\gamma_1$) decays on the order of seconds. Consistently, \reffig{fig2}a shows that lipid density relaxes with the rate of the fast mode. Since GUV flickering experiments usually operate at 1–1000 fps, they detect only the dynamics of the slow mode \cite{Faizi:2020}, and the experimentally measured bending rigidity is the thermodynamic one, $\kappa$, corresponding to relaxed lipid density. Furthermore, the bending mode relaxation rate is well-approximated by the value predicted from the planar bilayer theory, $\gamma_1\sim \kappa q^3/(4\eta)$ \cite{brochard.1975}.

For small liposomes with $R = 20$ nm, the system lies in the regime $\visrat_s \gg 1$, and its relaxation dynamics deviate significantly from the planar membrane theory, as shown in \reffig{fig1}b. In this case, the slow mode relaxes on a microsecond timescale, whereas the fast mode relaxes on a nanosecond timescale. Lipid density relaxation occurs on the slow-mode timescale, see \reffig{fig2}b; consequently, membrane undulations in the sub-microsecond regime are governed by the unrelaxed bending rigidity, $\Keff$. Because neutron spin echo (NSE) experiments typically probe nanosecond curvature fluctuations of SUVs \cite{Nagao:2017,Nagao:2023}, the slow mode is effectively frozen \cite{Watson:2010,Watson:2011}, and the measured correlation function reflects only the relaxation of the fast mode. Unlike the planar case, however, the decay rate is slower and asymptotically scales with the fourth power of the wavenumber.

\begin{figure}[h]
\includegraphics[width=0.5\columnwidth]{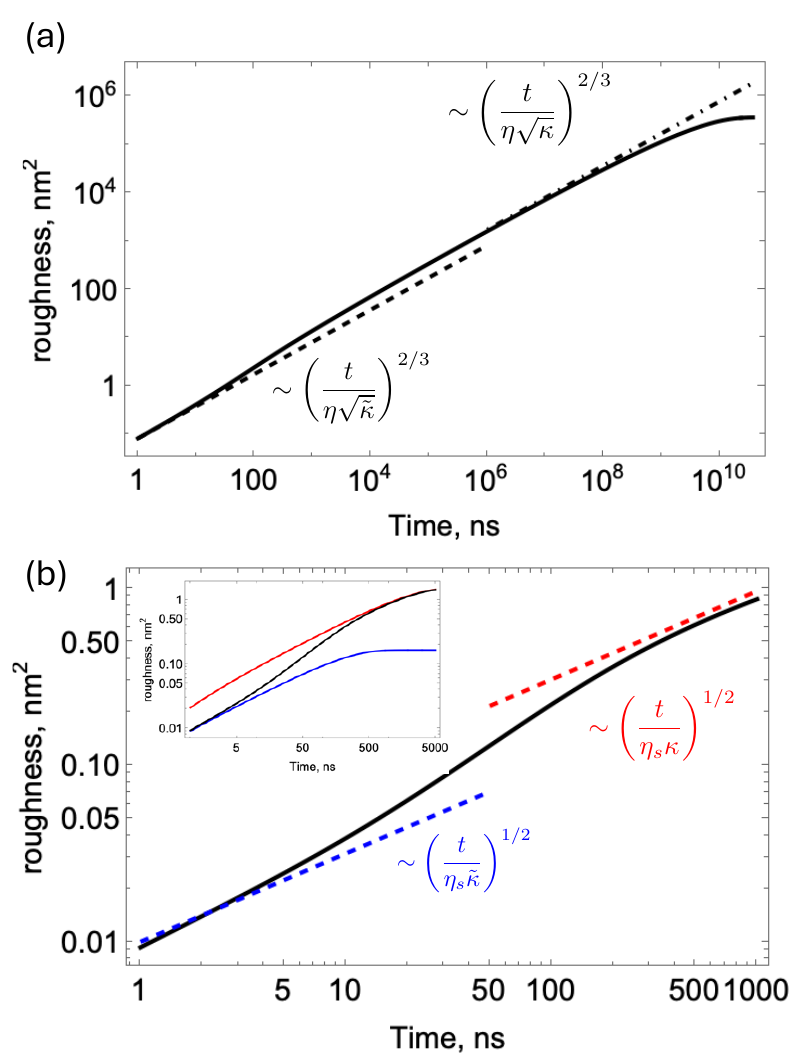}
\caption{\footnotesize{Single-point membrane MSD (roughness) of  vesicles. (a) and (b) correspond to the same parameters as in Figs. 2(a) (GUV) and 2(b) (SUV), respectively. The inset in (b) compares the MSD computed from the new theory (black) with that from the single–relaxation-time theory of Ref.~\cite{Faizi:2024}, given by \refeq{eqw}, using the relaxed bending rigidity $\kappa$ (red) and the unrelaxed bending rigidity $\Keff$ (blue).
\label{fig3}
}}
\end{figure}

\begin{figure}[h]
\includegraphics[width=0.5\columnwidth]{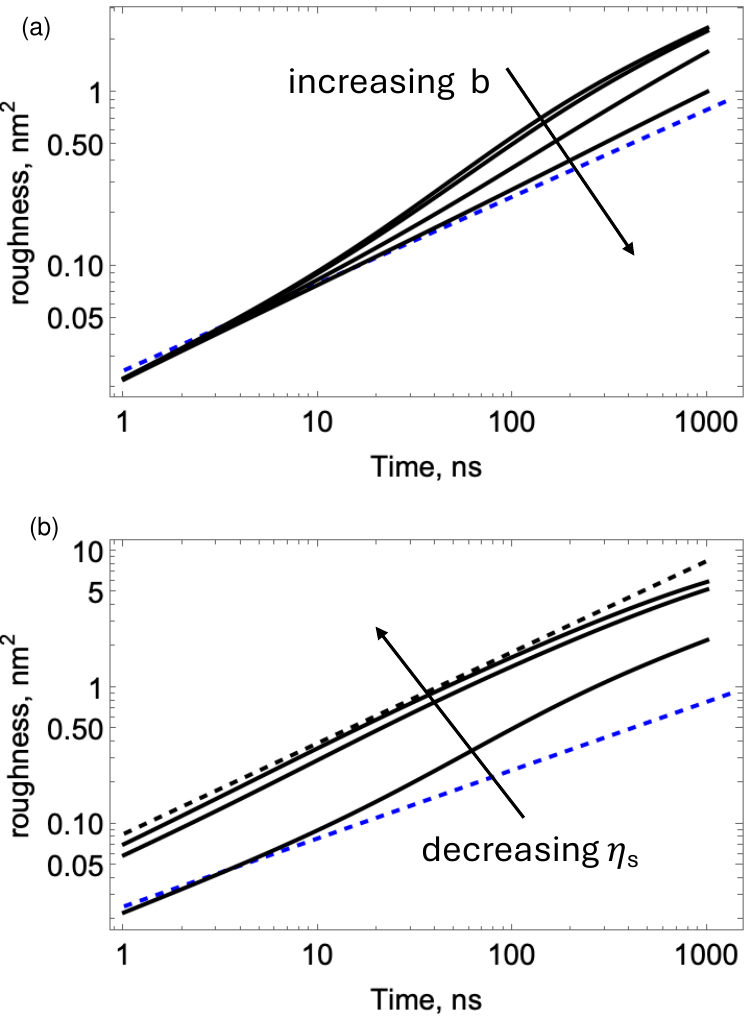}
\begin{picture}(0,0)(0,0)
\put(0,350){(a)}
\put(0,170){(b)}
\end{picture}
\caption{\footnotesize{Effect of bilayer slip (a) and membrane viscosity (b) on the single-point membrane MSD (roughness) of  a vesicle with radius 50nm. Material parameters same as in Fig. 2, except (a) bilayer slip  of increasing value: b=10$^4$, 10$^7$, to 10$^8$,10$^9$  N.m$^3$/s, and (b) membrane viscosity decreasing from $\eta_s=$$2.5\times 10^{-9}$ to $20\times 10^{-11}$  to $8\times 10^{-11}$ N.s/m(values reported from molecular dynamics simulations \cite{Fitzgerald:2023}). The black dashed line is the ZG asymptote, and the blue dashed line is the viscous asymptote \refeq{MSD-asymptotes}, both evaluated with the unrelaxed bending rigidity $\Keff$.
\label{fig5}
}}
\end{figure}

\reffig{fig3} illustrates the MSD behavior.
\reffig{fig3}a shows that for a GUV where the Saffman-Delbr\"uck length is smaller than the vesicle radius, and thus $\chi_s\ll1$, the MSD follows the ZG scaling for a planar membrane. Initially, the scaling is  with $\tilde\kappa$, since the lipid density is unrelaxed. As time progresses, the lipid density difference relaxes and the MSD approaches the ZG asymptotic scaling with $\kappa$. \colb{The crossover region is broad since different modes relax at different rate.}
However, for the small liposome with radius 20nm, and thus $\chi_s\gg 1$, see \reffig{fig3}b, no regions of clear power-law evolution exist. This is because of the fast dynamics -- the crossover times, $t_0$, $t^*$, and $t_R$ are too close {to one another; hence,} the separation between the crossover times is too small for the membrane relaxation to reach the asymptotic behavior. \colb{The inset of \reffig{fig3}b illustrates the crossover in the MSD evolution from a regime controlled by $\Keff$ to one governed by the relaxed bending rigidity $\kappa$. The crossover is sensitive to membrane viscosity and bilayer slip (see \reffig{fig5}). Increasing the intermonolayer friction slows the relaxation of the lipid density difference; accordingly, for $\chi_s \gg 1$ the dynamics can reach the viscous asymptotic regime governed by $\Keff$ (see \reffig{fig5}a). In contrast, decreasing the membrane viscosity (see \reffig{fig5}b) shifts the relaxation toward the asymptotic regime described by the ZG law.}

\begin{figure}[h]
\includegraphics[width=0.5\columnwidth]{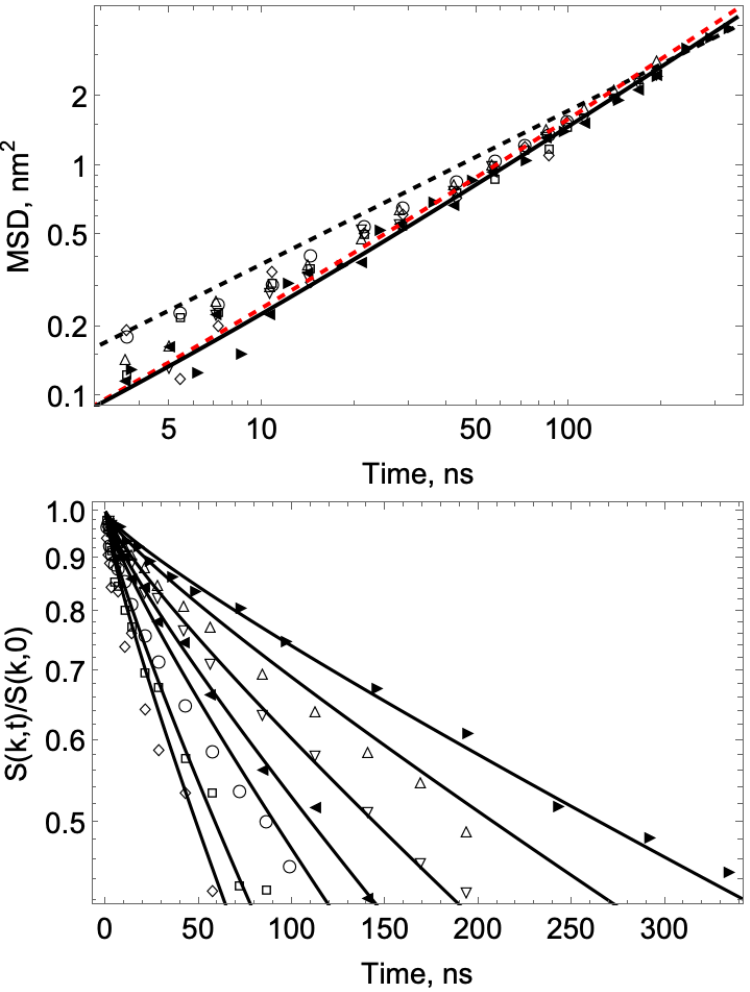}
\begin{picture}(0,0)(0,0)
\put(0,340){(a)}
\put(0,170){(b)}
\end{picture}
\caption{\footnotesize{ (a) MSD calculated from the experimental DSFs for POPC liposomes with radius  48nm
\cite{Granek:EPJE}. 
using membrane parameters $\kappa=25\kT$, $\K=122$ N/m$^2$, $d=1.45$nm\cite{Granek:EPJE}, and $b=5\times10^7$ N.s/m$^3$ \cite{Anthony:2022}. The black solid line corresponds to the new theory with membrane viscosity $7\times 10^{-10}$ N.s/m . {The black dashed line is the ZG-asymptote not corrected for liposome center-of-mass diffusion, \refeq{MSD-asymptotes}, using the ``unrelaxed'' bending rigidity  $\tilde\kappa$. The red dashed line is the ZG-asymptote with diffusional correction and effective bending rigidity $15\tilde\kappa$. (b) Comparison between the experimental and theoretical DSFs for scattering wavenumbers from 0.63  nm$^{-1}$ to 1.3 nm$^{-1}$, illustrating a good agreement for the long-time relaxation.}
}}
\label{fig4}
\end{figure}

What are the implications  of the new theory for the interpretation of the DSF measured by NSE experiments on liposomes? \col{The DSF for vesicle membranes was derived in Ref.\cite{Granek:EPJE}, and it was shown that it is well described by the MSD due to the polydispersity of the liposome suspension. 
}
\colb{\reffig{fig4} compares experimentally obtained values of the MSD—calculated from the DSFs {(neglecting scattering finite-size effects) as\cite{Granek:EPJE} $\ln\left(S(k,t)/S(k,0)\right)/(-\tfrac{1}{2}k^2)$ where $k$ is the scattering wavenumber}—with the theoretical prediction of \refeq{segment-MSD}, including the diffusional correction, $\langle \Delta f^2\rangle R^2 + 2Dt$ (the importance of properly correcting for diffusion has been discussed in Ref. \cite{Granek:EPJE,hoffmann2026describing}). 
Using the MD-reported viscosity \cite{Fitzgerald:2023} (interpolated to room temperature), $\eta_s \simeq 2\times10^{-10}$~N·s/m, versus the viscosity measured for GUVs \cite{Faizi:2022} (via the vesicle electrodeformation method), $\eta_s \simeq 1\times10^{-8}$~N·s/m, we find that the experimental data appear to be better described by an intermediate viscosity value. This suggests a possible scale dependence of the membrane 2D shear viscosity.  

\col{\reffig{fig4} illustrates that the data can also be fit using the ZG model, yielding results that may appear reasonable but are, in fact, physically inconsistent. 
The ZG model with the correction for liposome center-of-mass diffusion yields unphysically large bending rigidity ($\sim 15\tilde\kappa$). However, the uncorrected ZG model produces a bending rigidity that is misleadingly close to the expected “unrelaxed” value. The present theory, which consistently incorporates diffusion effects, provides an accurate description of the data 
without any assumptions for the bending rigidity being ``relaxed" or ''unrelaxed''. This also demonstrates that fitting NSE data with the ZG model without accounting for translational diffusion can be misleading, particularly for smaller vesicles. }

}

\section{Conclusions and open questions}
We theoretically analyze the spontaneous, thermally-driven shape fluctuations of a quasi-spherical vesicle made of a single component lipid bilayer. We derive an analytical
description of the dynamics of the shape and lipid density fluctuations with account of membrane viscosity thus extending the previous work by Miao et al. \cite{Miao:2002}. The new theory provides a unified description of the membrane dynamics in the broad time regime and wavelength spectrum spanning from cell-sized giant vesicles with radii of tens of microns down to the highly curved submicron  
liposomes.

We find that if the  Saffman-Delbr\"uck length $\eta_s/\eta$ is comparable to or bigger than the vesicle radius $R$, membrane viscosity significantly affects   the curvature fluctuations of liposomes compared to planar bilayers.
 Not only the membrane viscosity significantly reduces the relaxation of the bending mode, but also the asymptotic behavior of the decay rate and the DSF depart strongly from the Seifert-Langer and Zilman-Granek scalings, respectively. Furthermore, our analysis shows that curvature slows down the lipid {density} difference relaxation, and the DSF may not approach a stretched-exponential asymptotic behavior on the time scales of a typical NSE experiment (0.1-1000 ns). 
Accordingly, force-fitting the MSD with the ZG power law would lead to an overestimation of the unrelaxed bending rigidity $\tilde{\kappa}$.
\colb{The effect of membrane viscosity is significant despite the uncertainty in its value (macroscopic experiments report $\eta_m \sim 10^{-9}$~N·s/m \cite{Faizi:2022}, whereas molecular dynamics simulations \cite{Zgosrki:2019,Fitzgerald:2023} predict much lower viscosities.)}
 \colb{Only for very large liposomes and low-viscosity membranes do size and viscosity effects diminish, allowing curvature fluctuations to be well described by the ZG theory. A comprehensive comparison between theory and NSE experiments will be addressed in future work.}

We hope our findings will stimulate further studies into the effect of surface viscosity on the dynamics of membranes and other complex interfaces.

\section*{Conflicts of interest}
There are no conflicts to declare.

\section*{Data availability}
The codes to generate the Figures are available upon request from the authors.

\section*{Acknowledgments}
This research was supported by BSF Grant 2024173, and in part by 
the National Science Foundation under Grant NSF PHY-1748958. We thank Elizabeth Kelley, Ingo Hoffmann, and Michihiro Nagao for sharing the data for Figure 6 and for many helpful discussions.

\appendix

\section{Spherical harmonics}
\label{Harmonics}

The normalized spherical scalar harmonics are defined as
\begin{equation}
\label{normalized spherical harmonics}
 \textstyle{Y_{\jj m}\left(\theta,\varphi\right) =  \left[\frac{2\jj+1}{4\pi}\frac{(\jj-m)!}{(\jj+m)!}\right]^\half (-1)^m P_\jj^m(\cos\theta)e^{{\rm i}m\varphi},}
\end{equation}
where  $\rhat=\br/r$,  $(r, \theta,\varphi)$
are the spherical coordinates, and $P_\jj^m(\cos\theta)$ are the associated Legendre polynomials.
We define vector spherical harmonics  as
\begin{equation}
\label{vector harmonics}
\begin{split}
\bS_{\jj m0}&=r\nabla Y_{\jj m}=\frac{\partial Y_{\jj m}}{\partial \theta}  \that+\im m \frac{Y_{\jj m}}{\sin \theta}\phihat\,,\\
  \bS_{\jj m1}&=-\im \rhat \times \bS_{\jj m0}=-m\frac{Y_{\jj m}}{\sin \theta}\that-\im  \frac{\partial Y_{\jj m}}{\partial \theta}  \phihat\,,\\
   \bS_{\jj m2}&=\rhat Y_{\jj m}\,.\\
\end{split}
\end{equation}

\section{The equilibrium state}
\label{equil}

We choose $\rho_0$ as in Ref. \cite{Miao:2002}, which sets $\bar\phi$ to zero at equilibrium. If we assume the equilibrium density of each monolayer to be the same, it should be $\hat\rho_0=N^\pm/A^\pm$; $\rho_0=\hat\rho_0(1+d^2/R^2)$.

At equilibrium, if we assume unstressed sphere, $H=-1/R$, $A_\out=A_0\left(1+d/R\right)^2$ and $A_\ins=A_0\left(1-d/R\right)^2$, $\rho^\pm_0=\rho_0(1\mp 2d/R)$ (to a linear order in $d/R$)
\[
\phi_0=\frac{\rho_0^\out-\rho_0^\ins}{2\rho_0}=2d/R\,,\quad \bar\phi_0=\frac{\rho_0^\out+\rho_0^\ins}{2\rho_0}-1=0
\]
The energy of a spherical vesicle is 
\[
{\cal{H}}/A_0= 2\Keff/R^2 +\Km  \phi_0^2-4\Km (d/R)  \phi_0=2\kappa/R^2\,
\]
showing that the contributions from the elastic energy of the monolayers cancel.  

\section{Variation of the energy and membrane stresses}

Splitting the energy \refeq{FreeEnergy} into the contributions of each monolayer
\begin{equation}
\label{FreeEnergyM}
\begin{split}
\mathcal{H}^\pm
=&\frac{\Keff}{4} \int (2H)^{2}\dif A +\frac{\sigma_0}{2}\int \dif A+\frac{\K}{2} \int \left(\phi^\pm\right)^2\dif A\mp 2 d \K \int H\phi^\pm\dif A\,. 
 \end{split}
\end{equation}
To find the stresses, we consider the energy change upon variation of the interface $\delta \br=\Psi \bn+\Phi^i \bev_i$. We use the following relations \cite{deserno2015fluid}
\begin{equation}
\begin{split}
&\delta(\dif A)=\dif A\left(-2H\Psi+\nabla_i\Phi^i\right)\,,\quad \delta H=\left(2H^2-K_G\right)\Psi+\half\nabla^2 \Psi+\Phi^i\nabla_iH\\
&\delta(H\dif A)=\dif A\left(H\nabla_i\Phi^i-K_G\Psi+\half \nabla^2\Psi+\Phi^i\nabla_iH\right)
\end{split}
\end{equation}
Following \cite{Miao:2002}, we consider the number of molecules associated with any local area element $\dif A$ of a monolayer should be conserved under the shape variation of the monolayer
\begin{equation}
\delta\left(\rho^\pm \dif A\right)=0
\end{equation}
Thus we find the variation of the density fields
\begin{equation}
\delta \phi^\pm=-\left(1+\phi^\pm\right)\left(-2H\Psi+\nabla_i\Phi^i\right)
\end{equation}
Accordingly
\begin{equation}
\begin{split}
\delta \left(\left(\phi^\pm\right)^2\dif A\right)&=\left(\phi^\pm\right)^2\delta\left(\dif A\right)+\left(2\phi^\pm \delta \phi^\pm\right)\dif A=\left(-2H\Psi+\nabla_i\Phi^i\right)\left(-\left(\phi^\pm\right)^2-2 \phi^\pm\right)
\end{split}
\end{equation}
This amounts to renormalizing the monolayer's tension
\begin{equation}
\label{renormtens}
\sigma^\pm=\half \sigma_0-\Km \phi^\pm-\half \Km\left(\phi^\pm\right)^2
\end{equation}
The variation of the curvature-density coupling 
\begin{equation}
\delta\left(\phi^\pm H \dif A \right)=\phi^\pm \delta \left( H\dif A \right)+\delta\left(\phi^\pm \right) H \dif A
\end{equation}
\begin{equation}
\begin{split}
\delta\int \phi^\pm H \dif A=&\int \dif A\left(\left(\phi^\pm\left(2H^2-K_G\right)+2H^2+\half \nabla_s^2\phi^\pm\right)\Psi+\left(\phi^\pm+1\right)\nabla_s H\right) 
\end{split}
\end{equation}
\begin{equation}
\delta \int H^2 \dif A=\int \dif A\left(2H(H^2-K_G)+\nabla^2H \right)\Psi
\end{equation}

\begin{equation}
\delta \int \sigma \dif A =\int \dif A \sigma \left(-2H \Psi +\nabla_i\Phi^i\right)=-\int \dif A\left(2H\sigma +\nabla_s\sigma\right)
\end{equation}
after integration by parts and dropping the boundary terms.

Another way to find the variation is to consider a small deformation about  a sphere \cite{helfrich1986}, $r=R(r_0+f)$, $\phi=\phi_0+\psi$, $\bar\phi=\xi$.
 Using the shape function $F=r-R(r_0+f)$,
\begin{equation}
\bn=\frac{\nabla F}{|\nabla F|}=\frac{\rhat-\nabla f}{\sqrt{1+(\nabla f)^2}}
\end{equation}
The Jacobian, $\dif A=J\dif \Omega$, $J=r^2/(\bn\cdot \rhat)$ is 
\begin{equation}
J=R^2 (r_0+f)^2\sqrt{1+(\nabla_s f)^2}\approx R^2(r_0^2+2 f+f^2+\half (\nabla_s f)^2)+h.o.t.
\end{equation}

\begin{equation}
 \label{2H}   
 \begin{split}
 -2H=\nabla_s\cdot \bn&\approx \nabla_s\cdot\left(\left(\rhat-\frac{\nabla_sf}{r}\right)\left(1-\half \frac{(\nabla_sf)^2}{r^2}\right)\right)\\
 &=\nabla\cdot \rhat-\nabla_s\cdot\left(\frac{\nabla_sf}{r}\right)-\half \nabla_s\cdot\left(\frac{(\nabla_sf)^2}{r^2} \rhat\right)\\
 &=\frac{2}{r}-\frac{\nabla_s^2 f}{r^2}\\
 &=\frac{2}{R}\left(r_0-f-\half\nabla_s^2f+f^2+f\nabla_s^2f\right)
 \end{split}
\end{equation}

\begin{equation}
 \label{2H2}   
 (2H)^2=\frac{4}{R^2}\left(r_0^2-2f-\nabla_s^2f+3f^2+\frac{1}{4}\left(\nabla_s^2f\right)^2+3 f\nabla_s^2f\right)
\end{equation}

\begin{equation}
 \label{2H2J}   
 (2H)^2 J =4\left(1-\nabla_s^2f+f \nabla_s^2f+\half\left(\nabla_sf\right)^2+\frac{1}{4}\left(\nabla_s^2f\right)^2\right)
\end{equation}

\begin{equation}
 \label{2HJ}   
 (2H) J =-R\left(2r_0+2f-\nabla_s^2f+\left(\nabla_sf\right)^2\right)
\end{equation}

On a sphere $f=f_{\jj m} Y_{\jj m}$, $r_0=R(1-\frac{1}{4\pi} |f_{\jj m}|^2)$
Note that $\nabla f=\nabla_sf/r$ and $\nabla_s\cdot\left(r^{-2} Y_{\jj m} \rhat\right)=0$.
Expanding in spherical harmonics $\nabla_s f=\sqrt{j(j+1)} f_{jm} Y_{\jj m}$, $\nabla^2_s f=-\jj (\jj +1) f_{\jj m} Y_{\jj m}$

The $f_{00}$ amplitude is related to  the other amplitudes because of conservation of vesicle volume and it can be shown \citep{Seifert:1999, Vlahovska:2005} that
\begin{equation}
\begin{split}
&V=\frac{4\pi}{3} \left(1+\frac{f_{00}}{\sqrt{4 \pi}}\right)^3+\sum_{\jj \ge2} \sum_{m=-\jj}^{\jj} f_{\jj m}f_{\jj m}^*\,, \\
&f_{00}\approx -\frac{1}{\sqrt{4 \pi}}  \sum_{\jj \ge2} \sum_{m=-\jj}^{\jj} f_{\jj m}f_{\jj m}^*\,, 
\end{split}
\end{equation}
where $f^*_{\jj m}=(-1)^m f_{\jj -m}$.
Thus $V=4\pi/3+O(\eps^2)$ and at linear perturbation order, $O(\eps)$, volume is conserved.
The excess area $\Delta$ is also preserved to a leading order
\begin{equation}
\label{area constraint}
\begin{split}
\Delta=&A/R^2- 4\pi=\int \frac{\left(1+f\right)^2}{\hat\br\cdot\bn}\sin \theta \dif \theta\dif \varphi-4 \pi\\
&=\sum_{\jj m}\frac{\left(\jj+2\right)\left(\jj-1\right)}{2}f_{\jj m}f^*_{\jj m}+O(\eps^3)\,,
\end{split}
\end{equation}
where $\sum_{\jj m}\equiv  \sum_{\jj\ge2} \sum_{m=-\jj}^{\jj} $.
The outward normal vector to the vesicle surface defined by a shape function $F=r-1-f(\theta, \varphi, t)$ is
\begin{equation}
\bn=\frac{\nabla F}{|\nabla F|}=\rhat- \sum_{jm} \sqrt{\jj(\jj+1)} f_{\jj m}\bS_{\jj m0}+O(\eps^2)
\end{equation}
Accordingly, the mean curvature 
\begin{equation}
\label{Hcurv}
\begin{split}
H=-\half \nabla\cdot \bn&=
-1-\half \sum_{\jj m} \left(-2 +\jj(\jj+1)\right)f_{jm} Y_{\jj m}+O(\eps^2)
\end{split}
\end{equation}
where we used the fact that $\nabla\cdot \rhat =2/r=2(1-\eps f) +O(\eps^2)$ and  $\nabla^2_s Y_{\jj m}=-\jj(\jj+1)~Y_{\jj m}$ on a unit sphere.
The Gaussian curvature to a leading order in the deviation from a sphere is 
\begin{equation}
\label{Kcurv}
K_G=1+\sum_{\jj m} \left(-2 +\jj(\jj+1)\right)f_{jm} Y_{\jj m}
\end{equation}

The product of the vector spherical harmonics is recoupled as
\begin{equation}
\bS_{\jj_1m_10}\cdot\bS_{\jj_2m_20}=\chi\left(\jj_1,\jj_2,j\right)\zeta\left(\jj_1,\jj_2,\jj,m_1,m_2,m\right) Y_{\jj m}
\end{equation}
where
\begin{equation}
\label{chi}
\chi\left(\jj,\jj_1,\jj_2\right)=\frac{j\left(\jj+1\right)+\jj_1\left(\jj_1+1\right)-\jj_2\left(\jj_2+1\right)}{2 \left[\jj\left(\jj+1\right)\jj_1\left(\jj_1+1\right)\right]^{1/2}}\,,
\end{equation}
and the Clebsch-Gordan coefficient  is
\begin {equation}
\label{zeta}
\begin{split}
\zeta\left(\jj,\jj_1,\jj_2,m,m_1,m_2\right)=&{(-1)^{m_2}}
      \left[\frac{(2\jj+1)(2\jj_1+1)(2\jj_2+1)}{4\pi }\right]^{\half}\times
\left( \begin{array} {ccc}
     \jj& \jj_1& \jj_2\\
     0& 0& 0 
\end{array} \right )
      \left( \begin{array} {ccc}
     \jj& \jj_1& \jj_2\\
     m& m_1& -m_2 
\end{array} \right )\,.
\end{split}
\end{equation}
 $
\left( \begin{array} {ccc}
     \jj& \jj_1& \jj_2\\
     m& m_1& m_2 
\end{array} \right ) 
$
is the Wigner {\it 3j}\/-symbol. 
A special case for the {\it 3j}\/-symbol is
\begin {equation}
\label{3j-symbol =0:3}
\left( \begin{array} {ccc}
     \jj& \jj&0\\
     m& -m& 0 
\end{array} \right ) = \frac{(-1)^{\jj-m}}{\sqrt{2\jj+1}},
\end{equation}
Thus
The Jacobian expansion up to third order is 
\begin{equation}
\begin{split}
J=&\frac{\left(r_0+ f_{\jj m}Y_{\jj m}\right)^2}{\bn\cdot\hat\br}\\
&=r_0^2\left(1+\frac{1}{2}\eps^2\sum f_{\jj_1m_1}f_{\jj_2m_2}\chi\left(\jj_1,\jj_2,\jj_3\right)\zeta\left(\jj_1,\jj_2,\jj_3,m_1,m_2,m_3\right)Y_{\jj_3m_3}\right)\\
&+2r_0 f_{\jj m}Y_{\jj m}+ \sum f_{\jj_1m_1}f_{\jj_2m_2}\zeta\left(\jj_1,\jj_2,\jj_3,m_1,m_2,m_3\right) Y_{\jj_3m_3}+...
\end{split}
\end{equation}
Integration over a sphere requires
\begin{equation}
\int Y_{jm} Y_{j_1m_1}^*\dif \Omega= \delta_{jj_1}\delta_{mm_1}\quad\mbox{and} \quad \int Y_{jm}\dif \Omega=\sqrt{4 \pi}\delta_{j0}\delta_{m0}
\end{equation}
It  will kill all terms except  the ones for which $m_1=-m_2\equiv m$, $j_1=j_2\equiv j$, $j_3=0$, and $m_3=0$.

\begin{equation}
A/R^2=\int \frac{\left(r_0+\eps f_{\jj m}Y_{\jj m}\right)^2}{\bn\cdot\hat\br} d\Omega=4\pi r_0^2+\left[\frac{1}{2}f_{\jj m}^2 \jj (\jj +1)
+ f_{jm}^2\right]+...
\end{equation}
where we have taken into account that
\begin{equation}
\zeta\left(\jj,\jj,\jj,m,-m,0\right) =\frac{1}{\sqrt{4\pi}}
\end{equation}
The expansion for $r_0$ truncated at second order is given by 
\begin{equation}
r_0=1-\textstyle \frac{1}{4 \pi}f_{\jj m}^2
\end{equation}
and accordingly
\begin{equation}
r_0^2=1-2\textstyle \frac{1}{4 \pi}f_{\jj m}^2+..
\end{equation}
so from  $A/R^2=4\pi+\Delta$ we have for the excess area  
\begin{equation}
\Delta= f_{\jj m}^2\left[\frac{1}{2}j(j+1)-1\right]
\end{equation}
The change in the tension energy 
\[
\delta \left(\sigma_0\int dA\right)=\half R^2 \sigma_0  f_{\jj m}^2\left(\jj (\jj +1)-2\right)
\]
For the other energies
from \refeq{2H2} and the expression for the Jacobian
\begin{equation}
\begin{split}
\half \int (2H)^2dA&=8\pi+\left(-\jj(\jj+1)+\half \jj(\jj+1)+\frac{1}{4}\left(\jj(\jj+1)\right)^2 \right)f_{\jj m}^2\\
&=8\pi+\frac{1}{2}\jj(\jj+1)\left(\jj(\jj+1)-2\right)f_{\jj m}^2
\end{split}
\end{equation}

\begin{equation}
\begin{split}
\int (2H) \phi dA&=-R\left(\phi_0\left(8\pi +(-2+\jj(\jj+1)) f_{jm}^2\right)+\psi_{\jj m}f^*_{\jj m} (\jj(\jj+1)+2)\right)
\end{split}
\end{equation}

\begin{equation}
\begin{split}
\int (\phi)^2 dA&=R^2\left(\phi_0^2\left(4\pi +\half(\jj(\jj+1)-2) f_{jm}^2\right)+ 4 \phi_0\psi_{\jj m}f^*_{\jj m}+\psi_{\jj m}^2\right)
\end{split}
\end{equation}

The change in the energy due to the shape and density fluctuations
is

\begin{equation}
    \begin{split}
\delta E_\jj=&\half\Keff(\ell(\ell+1)-2)(\ell(\ell+1)+\bar \sigma ) f_{\ell m}^2 \\
&+\Km R^2 (\psi_{\ell m}^2+\xi^2_{\jj m})-2K_m d R \psi_{\ell m}f_{\ell m} (\ell(\ell+1)-2)
\end{split}
\end{equation}
where 
\[
\bar\sigma=\sigma_0+\alpha\phi_0^2-2\lambda \phi_0\,.
\]
If $\phi_0=2d/R$, then  $\alpha \phi_0^2=\lambda \phi_0$ and the tension $\bar\sigma =\sigma_0-\alpha \phi_0^2$.  This is consistent with the definition \refeq{renormtens}. When adding the tensions of the two monolayers, at equilibrium $\phi^-+\phi^-=0$, and the second term is $\frac{\Km}{4} (\phi^- -\phi^+)^2$. At equilibrium on a sphere $\phi^\pm=A_0(1\mp d/R)^2$.

\section{Fundamental set of velocity fields, tractions, and solution for the flow around a sphere}
\label{Ap:velocity basis}

The velocity basis functions that are regular at infinity are 
\begin{equation}
\label{vel basis -}
\begin{split}
\bu^-_{\jj m0}&={\textstyle\frac{1}{2}}r^{-j-2}\left((2-\jj)r^2+ \jj\right)\by_{\jj m0}+{\textstyle\frac{1}{2}}r^{-\jj-2}{\jj \left(\jj+1\right)}\left(
r^{2}-1\right) \by_{\jj m2} \, , \\
		\bu^-_{\jj m1}&=\textstyle r^{(-\jj-1)} \by_{\jj m1} \, , \\
		\bu^-_{\jj m2}&={\textstyle\frac{1}{2}}r^{-\jj-2}
(\textstyle\frac{\jj-2}{\jj+1})\left(1-r^{2}\right)\by_{\jj m0}+{\textstyle\frac{1}{2}}r^{-j}\left(\jj+(2-\jj)r^{2}\right)\by_{\jj m2}\,.
\end{split}
\end{equation}
The velocity basis functions that are regular at the origin are
\begin{equation}
\label{vel basis +}
\begin{split}
		\bu^+_{\jj m0}&= {\textstyle\frac{1}{2}}r^{\jj-1}\left(-(\jj+1)+(\jj
+3)r^2\right)\by_{\jj m0} -{\textstyle\frac{1}{2}}r^{\jj -1}{\jj \left(\jj+1\right)}\left
(1-r^2\right)\by_{\jj m2} \, , \\
\bu^+_{\jj m1}&=\textstyle r^\jj \by_{\jj m1} \, , \\
\bu^+_{\jj m2}&={\textstyle\frac{1}{2}}r^{\jj-1}(\textstyle\frac{3+\jj}
{j})\left(1-r^2\right)\by_{\jj m0}+{\textstyle\frac{1}{2}}r^{\jj-1}\left(\jj+3-(\jj+1)r^2  \right)\by_{\jj m2} \, .
\end{split}
\end{equation}
On a sphere $r=1$ these velocity fields reduce to the vector spherical harmonics defined by \refeq{vector harmonics}
\begin{equation}
\bu^{\pm}_{jmq}=\bS_{jmq}\,.
\end{equation}
Hence the continuity of normal velocity becomes simply
\begin{equation}
\label{velcont}
c^+_{jm2}=c^-_{jm2}
\end{equation}
The hydrodynamic tractions on a sphere due to the velocity fields \refeq{vel basis -} and \refeq{vel basis +} are
\begin{equation}
\label{HD trac:+}
\begin{split}
\tau^{\hd,\ins}_{\jj m0} &=\visrat\left(\textstyle{-(2\jj +1)c^+_{\jj m0}+\frac{3}{\jj} c^+_{\jj m2}}\right)\,,\\ 
\tau^{\hd,\ins}_{\jj m1} &=-\visrat (\jj -1)c^+_{\jj m1}\,,\\
\tau^{\hd,\ins}_{\jj m2} &=\visrat\left(\textstyle{3(\jj +1) c^+_{\jj m0}-\frac{3+\jj+2\jj^2}{\jj} c^+_{\jj m2}}\right)\,,
\end{split}
\end{equation}
\begin{equation}
\label{HD trac:-}
\begin{split}
\tau^{\hd,\out}_{\jj m0}& =\textstyle{ -(2\jj +1)c^-_{\jj m0}+\frac{3}{\jj +1}c^-_{\jj m2}}\,,\\
\tau^{\hd,\out}_{\jj m1}& =-(\jj +2)c^-_{\jj m1}\,,\\
\tau^{\hd,\out}_{\jj m2}& =\textstyle{3\jj c^-_{\jj m0}-\frac{4+3\jj+2\jj^2}{\jj +1} c^-_{\jj m2}}\,. 
\end{split}
 \end{equation}

\section{Evolution equations}
The full expressions for the evolution equations are
\begin{equation}
\label{evoleqsFull}
\bA\cdot \begin{pmatrix}  \dot{f}_{\jj m}\\ \dot{\psi}_{\jj m}\end{pmatrix}= -\bB\cdot \begin{pmatrix}  {f}_{\jj m}\\ \psi_{\jj m}\end{pmatrix}+\bF
\end{equation}
where $\bF$ is the thermal noise or external forcing (e.g., due to applied flow and electric field). 
\begin{subequations}
\label{Afull}
\begin{equation}
\begin{split}
A_{11}=&\frac{(4+3\jj^2+2\jj^3)+(-5+3\jj^2+2\jj^3)\visrat+4(-2+\jj+\jj^2)\visrat_s}{\jj(\jj+1)}
\end{split}
\end{equation}
\begin{equation}
\begin{split}
A_{12}=&\frac{ \left((2+\jj)(1-\phi_0)-(\jj-1)\visrat(1+\phi_0)-2\left(2\jj(\jj+1)\visrat_d +(-2+\jj+\jj^2)\visrat_s\right)\phi_0 \right)}{\jj(\jj+1)\left(1-\phi_0^2\right)}
\end{split}
\end{equation}
\begin{equation}
\begin{split}
A_{21}=&\frac{(2 + \jj) ( 1-\phi_0) -2 (-2 + \jj + \jj^2) \visrat_s\phi_0 -( \jj-1) \visrat(1 + \phi_0)}{\jj(\jj+1)}
\end{split}
\end{equation}
\begin{equation}
\begin{split}
 A_{22}=\frac{ \left(4\beta+(2\jj+1)((\phi_0-1)^2+\visrat (\phi_0+1)^2)+(\jj(\jj+1)\visrat_d +(-2+\jj+\jj^2)\visrat_s)\left(\phi_0^2+1\right)\right)}{{\jj(\jj+1)\left(1-\phi_0^2\right)}}
\end{split}
\end{equation}
\end{subequations}
\begin{equation}
\label{mB}
\begin{split}
B_{11}=&(\jj-1)(\jj+2)\left(\jj(\jj+1)+\bar\sigma\right)\,\\
B_{12}=&-(\jj-1)(\jj+2)\lambda\,,\\
B_{21}=&-(\jj-1)(\jj+2)\lambda \left(1-\phi_0^2\right),\,\\
B_{22}=&2\alpha\left(1-\phi_0\right)^2\,.
\end{split}  
\end{equation}
The tension $\bar\sigma=\sigma_0-\alpha \phi_0^2$.

\begin{figure}[h]
\includegraphics[width=\linewidth]{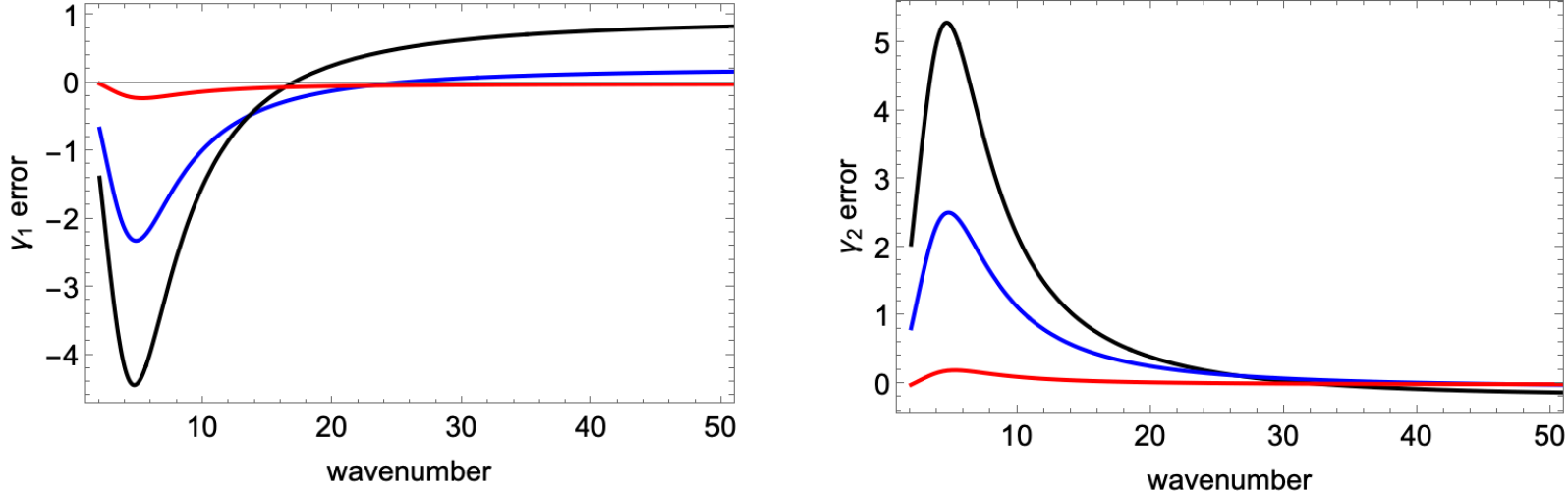}
\caption{\footnotesize{Relative error (in percents) in the relaxation rates computed using the approximate, \refeq{mM} and \refeq{mE},  and full expressions, \refeq{Afull} and \refeq{mB}, as a function of the wave number $q$ for liposomes with radii $R=$ 50nm (black, 100nm (blue), and 1$\mu$m (red).
\label{fig6}
}}
\end{figure}

The diagonal elements of the matrix {\bf{A}} are much larger than the off-diagonal ones and the latter can be approximated by zero; for the parameters in Ref.\cite{Watson:2011} the error in the relaxation rates is size-dependent (it increases with liposome radius),  but below 0.5\% for modes above 20, see Figure \ref{fig6}. The corresponding error in the roughness for the 50nm, 100nm and 1$\mu$m liposomes is below 2\%, 1\% and 0.01\%, respectively.

In the case of  $\phi_0= 0$ and  $\visrat=1$, the asymptotic behavior of $\bA$ for high $\ell\gg \visrat_s \gg 1$ yields the Seifert-Langer theory for a tensionless membrane $\sigma_0=0$\cite{Seifert-Langer:1993,Fournier:2015}
\[
\begin{split}
A_{11}&\approx 4 \jj \,,\quad A_{22}\approx \jj^{-2}(4\beta+4\jj+\jj^2(\chi_s+\chi_d)) \,,\quad
A_{12}=A_{21}\approx 0\\
E_{11}&\approx\jj^4\,,\quad E_{12}=E_{21}\approx -\jj^2\lambda\,,\quad \quad E_{22}\approx 2\alpha\,.
\end{split}
\]

\bibliographystyle{unsrt}

\end{document}